\documentclass{emulateapj}
\usepackage{apjfonts,lscape}

\journalinfo{To appear in the Astrophysical Journal Supplements}
\slugcomment{}

\shorttitle{MUSYC:  Images and Catalogs of E-HDFS}
\shortauthors{Gawiser et al.}

\begin{document}

\title{The Multiwavelength Survey by Yale-Chile (MUSYC): Survey Design and 
Deep Public UBVRI\lowercase{z$'$} Images and Catalogs 
of the Extended Hubble Deep Field 
South\footnotemark[1]
\vspace{0.2 in}
}

\author{Eric Gawiser\footnotemark[1,2,3,4], 
Pieter G. van Dokkum\footnotemark[2,3],
David Herrera\footnotemark[2,3,5],
Jos\'e Maza\footnotemark[1],
Francisco J. Castander\footnotemark[1,6],
Leopoldo Infante\footnotemark[7],
Paulina Lira\footnotemark[1], 
Ryan Quadri\footnotemark[2],
Ruth Toner\footnotemark[3],
Ezequiel Treister\footnotemark[1,2,3],
C. Megan Urry\footnotemark[3,5], 
Martin Altmann\footnotemark[1],
Roberto Assef\footnotemark[7],
Daniel Christlein\footnotemark[1,2,3],
Paolo S. Coppi\footnotemark[2,3,5],
Mar\'{\i}a Fernanda Dur\'an\footnotemark[1],
Marijn Franx\footnotemark[8],
Gaspar Galaz\footnotemark[7], 
Leonor Huerta\footnotemark[1],
Charles Liu\footnotemark[9], 
Sebasti\'{a}n L\'{o}pez\footnotemark[1], 
Ren\'{e} M\'{e}ndez\footnotemark[1],
David C. Moore\footnotemark[2], 
M\'{o}nica Rubio\footnotemark[1], 
Mar\'{\i}a Teresa Ruiz\footnotemark[1],
Sune Toft\footnotemark[2,3],
Sukyoung K. Yi\footnotemark[10]
}

\affiliation{}

\footnotetext[1]{Departamento de Astronom\'{\i}a, Universidad de Chile, Casilla 36-D, Santiago, Chile.}
\footnotetext[2]{Department of Astronomy, Yale University, PO Box 208101, New Haven, CT  06520.}
\footnotetext[3]{Yale Center for Astronomy \& Astrophysics, Yale University, 
P.O. Box 208121, New Haven, CT 06520.}
\footnotetext[4]{NSF Astronomy and Astrophysics Postdoctoral Fellow.}
\footnotetext[5]{Department of Physics, Yale University, New Haven, CT, PO Box 208121, New Haven, CT  06520.}
\footnotetext[6]{Institut d'Estudis Espacials de Catalunya/CSIC,Gran Capit\`a 2-4, E-08034 Barcelona, Spain.}
\footnotetext[7]{Departamento de Astronom\'{\i}a y Astrof\'{\i}sica, 
Pontificia Universidad Cat\'olica de Chile, Casilla 306, 22 Santiago, Chile.}
\footnotetext[8]{Leiden Observatory, Postbus 9513, NL-2300 RA Leiden, Netherlands.}
\footnotetext[9]{Astrophysical Observatory, City University of New York, 
College of Staten Island, 2800 Victory Blvd., Bldg. 1N-232, 
Staten Island, NY  10314.}
\footnotetext[10]{University of Oxford, Astrophysics, Keble Road, Oxford OX1 3RH}

\footnotetext[1]{Based 
on observations obtained at Cerro Tololo Inter-American 
Observatory, a division of the National Optical Astronomy Observatories, 
which is operated by the Association of Universities for Research in 
Astronomy, Inc. under cooperative agreement with the National Science 
Foundation.}

\email{gawiser@astro.yale.edu}

\begin{abstract} 
We present $UBVRIz'$ optical images taken with CTIO4m+MOSAIC of the 
0.32 deg$^2$ 
Extended Hubble Deep Field South.  
This is one of four fields comprising the 
MUSYC survey, which is optimized for the 
study of galaxies at $z=3$, AGN demographics, and Galactic structure.  
Our methods used for astrometric calibration, weighted 
image combination, and photometric calibration in AB magnitudes 
are described.  
  We calculate corrected aperture 
photometry and its uncertainties and find through tests that 
these provide a significant improvement upon standard techniques.  
Our photometric catalog of 62968 objects is complete to a total magnitude 
of $R_{\rm AB}=25$, 
with $R$-band counts consistent with results from the literature. 
We select $z\simeq 3$ Lyman break galaxy (LBG) 
candidates from their $UVR$ colors 
and find a sky surface density of 1.4 arcmin$^{-2}$ and an 
angular correlation function $w(\theta) = 2.3\pm 1.0 \theta^{-0.8}$, 
consistent with previous findings  
that high-redshift Lyman break galaxies 
reside in massive dark matter halos.
Our images and catalogs are available at 
\url{http://www.astro.yale.edu/MUSYC}.
\end{abstract}


\section{INTRODUCTION}
\label{sec:intro}

\setcounter{footnote}{1}

The study of galaxy formation and evolution requires detailed information 
about statistically significant samples of dim objects.  This, in turn, 
requires deep imaging and spectroscopy over wide areas of the sky.  In 
pursuit of these data, several wide-deep surveys are now underway.  Those 
covering several square degrees or more 
either lack spectroscopic follow-up    
(e.g. NOAO DWFS, \citealt{jannuzid99}, and ODTS, 
\citealt{macdonaldetal04})   
or are restricted to the study 
of objects at 
$z \la 1$ (except for quasars)  
by their imaging depth e.g. 
the Sloan Digital Sky Survey 
(\citealt{yorketal00} and 
\citealt{abazajianetal05}), 
the VIRMOS-VLT Deep 
Survey (\citealt{lefevreetal04}, \citealt{radovichetal04})
and DEEP2 \citep{davisetal03}.   
Other surveys target the high-redshift universe with deep HST imaging 
over fractions of a square degree i.e. the Hubble 
Deep Fields \citep{williamsetal96,williamsetal00}, the Hubble Ultra Deep 
Field (HUDF), 
GOODS \citep{giavaliscoetal04, dickinsonetal04}, 
and GEMS \citep{rixetal04}.    
Spectroscopic coverage is feasible over these areas but is presently 
unable to probe deeper than $R\simeq 25$, making the added imaging 
depth useful only for morphological studies and photometric redshifts.  

The Multiwavelength Survey by Yale-Chile (MUSYC)
probes the intermediate regime of 
a square degree of sky
to the spectroscopic limit of $R\simeq 25$.  
Section 2 describes 
the design of our survey.  \S 3 reports our imaging observations 
for the extended Hubble Deep Field South. \S 4 describes our imaging 
reduction, and \S 5 covers our photometric calibration and photometry.  
\S 6 gives our results for $R$-band number 
counts and the sky density and angular clustering of 
$UVR$-selected Lyman break galaxies. \S 7 concludes.  
Our analyses assume a standard $\Lambda$CDM cosmology with 
$\Omega_m$=0.27,$\Omega_\Lambda=0.73$, and $H_0$=70km/s/Mpc.  

\section{SURVEY DESIGN}
\label{sec:sample}

The Multiwavelength Survey by Yale-Chile 
(see \url{http://www.astro.yale.edu/MUSYC}) 
is designed to provide a fair sample of the universe for the study of 
the formation and evolution of galaxies and their central black holes.  
The core of the survey is a deep imaging campaign in 
optical and near-infrared passbands of four carefully selected 
$30'\times30'$ fields.  
MUSYC is unique for its 
combination of depth and total area, 
for additional coverage at X-ray, UV, mid- and far-infrared
wavelengths and 
for providing the $UBVRIz'JHK$ photometry 
needed for high-quality photometric redshifts over a square-degree 
of sky.  
The primary goal is to study the properties 
and interrelations of galaxies at a single epoch corresponding to 
redshift $\sim3$, using a range of selection techniques.
  We chose to use the $UBVRIz'$ filter set in the optical in order to obtain 
six nearly-independent flux measurements with the 
broadest possible wavelength coverage.    

  Lyman break galaxies at $z\simeq 3$ 
are selected  
through their dropout in $U$-band images 
combined with blue continuua in $BVRIz'$ 
($\lambda > 1216${\AA} in the rest-frame)
typical of recent star formation \citep{steideletal96, steideletal99, 
steideletal03}.
Imaging depths 
of $U$,$B$,$V$,$R \simeq 26$ were chosen 
to detect the LBGs, whose luminosity 
function has a characteristic magnitude of $m_* = 24.5$ in $R_{\mathrm AB}$, 
and to find their Lyman break decrement in 
the $U$ filter via colors $(U-V)_{\mathrm AB} > 1.2$.    

Lyman $\alpha$ emitters 
at $z\simeq 3$ are selected through additional deep narrow-band imaging using 
a 50{\AA} fwhm filter centered at 5000{\AA}.  These objects can be detected 
in narrow-band imaging and spectroscopy 
by their emission lines, 
allowing us to 
probe to much dimmer continuum magnitudes than possible for Lyman 
break galaxies.  

It has recently become clear that optical selection methods do not 
provide a full census of the galaxy population at $z\sim3$, as they 
miss objects which are faint in the rest-frame ultraviolet 
\citep{franxetal03,daddietal04a}.  With this in mind, MUSYC has 
a comprehensive near-infrared imaging campaign.  The NIR 
imaging comprises two components:
a wide survey
covering the full 
square degree 
and a deep survey
of the central $10'\times 10'$ of each field. This division between
deep and wide was chosen because of the $10.5'$ field-of-view of
the ISPI near-infrared camera on the CTIO 4m telescope.
The $5\sigma$ point source sensitivities of the wide and deep
components are $K_{s,AB}=22.0$ and $K_{s,AB}=23.3$ respectively. 
NIR imaging over the full survey area provides a critical complement 
to optical imaging for breaking degeneracies in photometric redshifts 
and modeling star formation histories.  Deeper $JHK_s$ imaging 
over $10'\times 10'$ subfields opens up an additional window 
into the $z\simeq3$ universe 
as the $J-K$ 
selection technique \citep{franxetal03, vandokkumetal03, vandokkumetal04} 
will be used to 
find evolved optically-red galaxies at $2<z<4$ through their rest-frame 
Balmer/4000{\AA} break.  

Extensive follow-up spectroscopy is being conducted over the 
square degree.  
 A subset of the color-selected Lyman break galaxy candidates will turn out 
to be AGN based on broad- or narrow-line emission features seen in 
follow-up spectroscopy \citep{steideletal02}.  
Damped Lyman $\alpha$ Absorption systems 
\citep{wolfeetal86} at $z>2.3$, which comprise the neutral gas 
reservoir needed to form most of the stars in the universe
\citep{wolfegp05}, 
will be searched for in the spectra of 
the brightest color-selected LBG/AGN candidates (typically quasars 
at $z\sim3$).

In addition to the optical and near-infrared, imaging campaigns 
at other wavelengths and follow-up spectroscopy are integral parts 
of MUSYC.  
X-ray selection will be used to study AGN demographics
over the full range of accessible redshifts, $0<z<6$, 
\citep[see][]{liraetal04} 
with Spitzer imaging 
used to detect optically- and X-ray-obscured AGN
\citep{treisteretal04b,lacyetal04}.  
This also allows a census of accreting black holes at $z\simeq 3$ 
in the same fields to study correlations between black hole accretion 
and galaxy properties at this epoch.  

Future epochs of optical 
imaging will be used to conduct a proper motion survey to 
find white dwarfs and brown dwarfs in order to study Galactic structure 
and the local Initial Mass Function
\citep[see][]{altmannetal04}; the additional epochs will also 
enable a variability study of AGN.  

The four 
survey fields (see Table~\ref{tab:fields}) were  
chosen to have 
extremely low reddening, HI column density \citep{bursteinh78}, 
and 100$\mu$m dust emission \citep{schlegelfd98}  
in order to 
facilitate satellite coverage with Spitzer, HST, Chandra,  
and XMM,  
to take advantage of existing multiwavelength data and to enable flexible
scheduling of observing time during the year.
Additionally, each field satisfies 
all of the following criteria:
minimal bright foreground sources in the optical and radio, 
high Galactic latitude ($|b|>30$) to reduce stellar density, 
and accessibility from observatories located in Chile.  
The survey fields will be a natural
choice for future observations with ALMA.

The remainder of this paper 
describes our optical images and catalog of E-HDFS.\footnote{The 
data presented here are 
available at\\ \url{http://www.astro.yale.edu/MUSYC} } 
The techniques used for data reduction and photometry are the same 
as those used for the analysis of the other three fields.  Optical 
imaging from the full survey will be reported in E.~Gawiser et al. (2005,  
in preparation).  The near-infrared data will be discussed in 
R.~Quadri et al. (2005, in preparation).  
The E-HDFS has deep public space-based 
observations at UV, optical, near-infrared, and far-infrared 
wavelengths.  The HDFS itself covers 
a small $\sim2.5'\times2.5'$ central region with WFPC2 plus STIS and 
NICMOS regions, with 
deep ground-based JHK coverage of the WFPC2 region available from the FIRES 
survey \citep{labbeetal03}.    
Spitzer IRAC and MIPS coverage of the central 
$5'\times15'$ is being performed in GTO time. 
The extended area around HDFS has previously been imaged by 
\citet{palunasetal00} and \citet{teplitzetal01} to a depth sufficient for the 
study of galaxies at $z<1$.  These images were made public and 
combined with deep H-band images by the Las Campanas Infrared Survey 
(LCIRS, \citealt{chenetal02}) to study red galaxies out to $z\simeq 1.5$.  
Our survey goes about one magnitude 
deeper in UBVRI$z'$ to probe the $z=3$ universe.  
Our Extended Hubble Deep Field South (E-HDFS) field center 
(see Table~\ref{tab:fields}) 
was chosen to keep a bright star (m=6.8) which lies just North of 
the WFPC2 field off of the CTIO+MOSAIC detectors.    
LCIRS covered an H-shape centered on WFPC2 and thereby 
provides public $H$-band coverage of roughly half of our E-HDFS 
field.

\section{OBSERVATIONS}
\label{sec:observations}

\begin{figure}
\epsscale{1.0}
\plotone{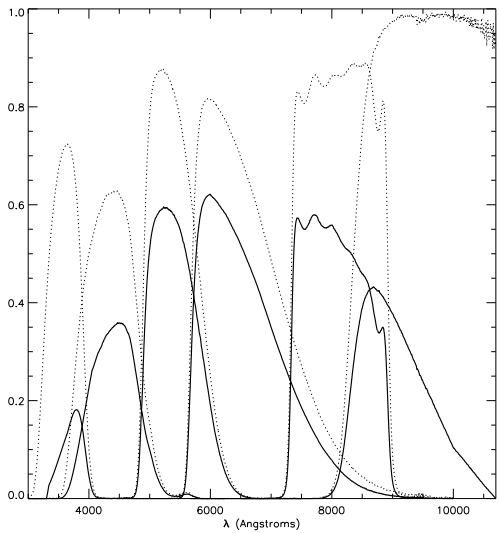}
\caption{
MOSAIC II $UBVRIz'$ filter set, starting with $U$ at left.  Dotted lines 
show filter throughput, and solid lines show total system throughput 
after multiplying by CCD quantum efficiency and 
atmospheric transmission at one 
airmass.
\label{fig:filters}
}
\end{figure}

Optical images of E-HDFS were taken on the nights of 2002 October 
6,8,10,12 and 2003 May 26,27,28 using the 8192$\times$8192 pixel 
MOSAIC II camera consisting of 8 2048$\times$4096 CCDs, each 
with 2 amplifiers, on the Blanco 4m telescope at CTIO.  
Afternoon calibrations were obtained, including zero-second exposures 
to trace the readout bias pattern, and domeflats in each filter to 
be observed except for $U$ for which the counts from the dome lamps 
were insufficient.  Twilight flats were therefore obtained 
for the $U$ filter.
Dark exposures of comparable length to the imaging 
observations were obtained, but due to the negligible dark 
current they were not used for data reduction.
A standard dither pattern was used to 
fill in gaps between the 8 CCDs.  The pixel scale 
is 0.267$''$/pixel, leading to coverage spanning $37'\times 37'$ of sky 
with each pointing.  
Figure \ref{fig:filters} shows the filter response curves and 
their multiplication with the CCD quantum efficiency and atmospheric 
transmission at one airmass.  
Table \ref{tab:obs} gives details of the exposure times 
in each filter for each run along with the approximate average 
seeing measured in the raw images during observations.
The seven nights were mostly cloudless, but moderate clouds 
affected some of the $V$ imaging of E-HDFS on 2002 October 10,12 and 
some of the $R$ imaging on 2003 May 28; our reduction methods 
described below allow these images to be used without biasing 
the photometry.    
Photometric standard fields from \citet{landolt92} large enough 
to cover the full MOSAIC II field of view were 
observed each night, and the nights of 2002 October 6 and 2003 May 26 
proved photometric.

\section{DATA REDUCTION}
\label{sec:reduction}

$UBVRIz'$ images from CTIO4m+MOSAIC II 
were reduced using the MSCRED and MSCDB packages in IRAF 
v.2.12\footnote{IRAF is distributed by the National Optical Astronomy 
Observatory, which is operated by the Association of Universities 
for Research in Astronomy, Inc., under cooperative agreement with the 
National Science Foundation.}
following the NOAO Deep Wide Field Survey 
cookbook v.7.02.\footnote{The NDWFS cookbook can be found at\\ 
\url{http://www.noao.edu/noao/noaodeep/ReductionOpt/frames.html}}
We used custom software to 
work around a few difficulties in these 
packages, as described below.  

A composite zero image is subtracted from each raw image to 
remove the amplifier bias level and pattern. The resulting image is 
then ``flat-fielded'' by dividing by a composite domeflat, or a composite 
twilight flat in the case of $U$-band.  A {\it superskyflat}
for each filter is made 
by combining all of the flat-fielded, unregistered images taken in each 
filter each night with rejection used to remove 
sources.  
Our 2$'$ amplitude dither pattern was designed 
to eliminate the wings of bright sources from the {\it superskyflat}.  
Each flat-fielded 
image is then divided by the appropriate {\it superskyflat}, which 
offers an estimate of the pixel-by-pixel response to 
the spectrum of the night sky with sufficient counts to 
achieve 1\% precision per pixel.  Because the {\it superskyflat} was produced 
using flat-fielded images, dividing by it serves to remove the 
original domeflat (twilight flat) from the reduction process.  The real 
influence of the original flat-fielding is to remove the illumination 
pattern and gross pixel-to-pixel variations before looking for cosmic 
rays and bright objects to reject in making the {\it superskyflat}.  
Using the {\it superskyflat} to correct for the pixel response
 is considered preferable 
to using the domeflat because the CCD response to the spectrum of the 
dome lamp (or twilight sky) can have significant systematic 
differences from its response to the spectrum of 
the night sky.  For background-limited photometry, 
this is an important effect.  

We then find an astrometric solution for each image, starting 
with fiducial WCS headers provided for each fits extension of the 
 raw images and comparing the claimed positions with the known 
positions of stars in the USNO-B catalog using the MSCRED routine 
{\it msccmatch}.  
 We found 
{\it msccmatch} to be finicky; for some runs the fiducial 
headers were too inaccurate to be corrected with the maximum second-order 
terms used by the MSCRED package.  An iterative, non-interactive procedure 
of calling {\it msccmatch} multiple times proved sufficient.   
The final rms
astrometric errors are between 0.2$''$ and 0.3$''$ in each 
image, which is consistent with the uncertainties on individual USNO-B 
stars despite having fit many more stars than free parameters, and 
the actual solution should be better than 0.2$''$ rms.  

To perform 
aperture photometry later on, 
our final images are transformed to have a common 
pixel scale and tangent plane projection point.  This 
is accomplished by projecting each processed image (after bias-subtraction, 
flat-fielding, {\it superskyflat}-fielding, and correction of 
WCS header information) onto the tangent 
plane of a  common reference image using {\it mscimage}.
The reprojection performed by {\it mscimage} 
should be performed with flux conservation set to {\it no} the first time 
but then set to {\it yes} for any further reprojections.  This is because 
the pixel scales in raw MOSAIC II images are a function of radius from the 
field center with several percent variation from center to corner of 
the field.  The process of flat-fielding removes the true illumination 
pattern caused by more photons falling in larger pixels leading to an 
illusion of flat background counts despite the variation in pixel 
scales.  When 
{\it mscimage} is used for astrometric projection into a uniformly 
sized grid of pixels, flux conservation would re-introduce the 
illumination pattern, thereby causing a variable photometric 
zeropoint across the image.  
Turning off flux conservation instead produces an average 
of the values of initial pixels neighboring each final pixel location
that offsets the error introduced by flat-fielding.

We make three different stacked ``final'' images for each filter.  
Unweighted versions are 
made first in order to 
determine the weights for the other two versions.  
The versions denoted by {\it xs} were combined using weights
 optimized for surface brightness, 
which is the method used by the NDWFS cookbook.
The versions denoted by {\it ps} were made using 
weights optimized for point sources; 
details can be found in the Appendix. 

We synthesize a $BVR$\_{\it ps} 
image to use as our photometric detection image by 
adding the $B$, $V$, and $R$ {\it ps} 
stacks using point-source-optimized 
weights derived from the signal and noise statistics 
of these stacks.
We decided not to add $U$ to $BVR$ although it is also very  deep because 
some low-redshift objects have different morphologies in $U$, 
the seeing is typically worse in $U$, 
and 
objects at $z\simeq 3$ will be nearly invisible in $U$ so adding it to 
$BVR$ would just add noise for these objects.  
Note that for similar reasons 
objects that drop out in $B$ and $V$  are somewhat less likely 
to be detected in our $BVR$ image than in a $VR$ or $R$ image alone.  
We prefer the $BVR$ approach to the creation of a $\chi^2$ image 
advocated by \citet{szalaycs99} because it allows us to measure 
object morphological parameters directly from the same image used 
for object detection, which is not recommended in the latter approach.
Moreover, the optimally-weighted $BVR$ combination reduces the 
emphasis on single-filter outliers employed by the $\chi^2$ approach and 
avoids its flaw of treating both negative and positive sky fluctuations  
as evidence of an object.  

We trim the final $BVR$\_{\it ps} image to the maximum size region that 
has nearly uniform signal-to-noise ratio   
in all three input images $B$, $V$, and $R$.  
For E-HDFS, the trimmed image is 7395$\times$7749 pixels or $33'\times34.5'$.
Then the IRAF task 
{\it imalign} is used to shift and trim the other images to match, giving 
our images the common origin and size required for aperture photometry 
with SExtractor.  
The trimmed images are normalized to an effective exposure time 
of one second.
After the pipeline reduction, the background is flat to better than 
1\% in all filters.  The remaining low level, large scale fluctuations 
were 
subtracted 
using SExtractor \citep{bertina96} 
with mesh size BACK\_SIZE=64 and median filter 
BACK\_FILTERSIZE set to 6 mesh units.

\begin{figure}
\epsscale{1.2}
\plotone{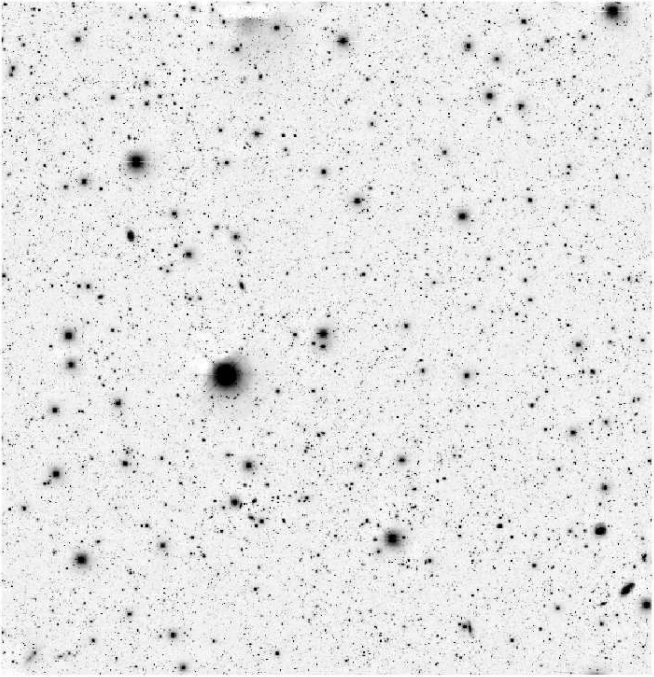}
\caption{
$33'\times34.5'$ combined $BVR$ image of E-HDFS used for object detection.
The mild background feature at the top just left of center  
is caused by the 
reflection of a magnitude 6.8 star just outside the field.    
\label{fig:BVR}
}
\end{figure}

\begin{figure}
\epsscale{1.2}
\plotone{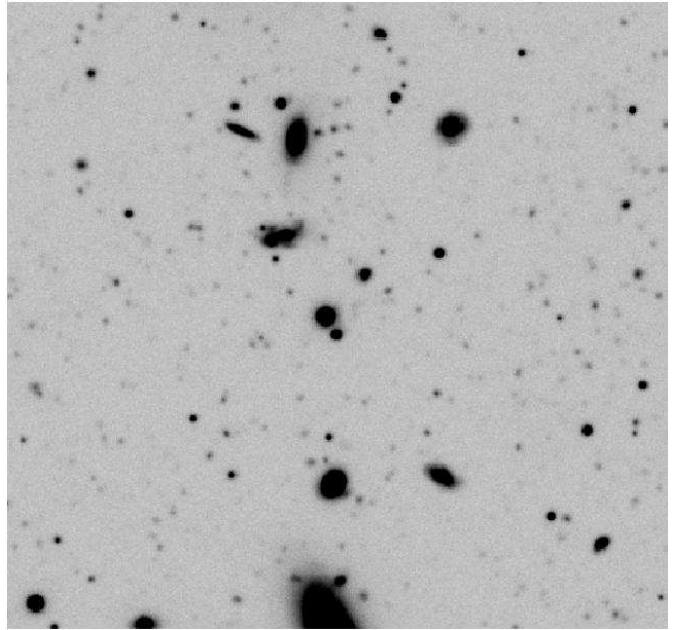}
\caption{
$2'\times2'$ zoom of $BVR$ detection 
image.  The seeing in the image is 0.95$''$ fwhm.
\label{fig:zoom}
}
\end{figure}

The resulting final $BVR$ image is shown in Figure \ref{fig:BVR} with 
a magnified image of a small section shown in 
Fig. \ref{fig:zoom}.  
New FITS 
headers have been
 added to these final images indicating input information 
for SExtractor photometry runs and other routines, specifically:  \\
SATUR\_LEVEL, the empirically determined saturation level in each 
image, which is usually a factor of a few less than the apparent 
saturation level of the brightest stars, \\
SEEING\_FWHM, the mode of the seeing for a set of bright unsaturated stars,  
\\
MAG\_ZEROPOINT, the AB magnitude of an object that gives 
1 count per second derived from photometric 
calibration as described below, \\
FLUX\_ZEROPOINT, the flux in $\mu$Jy corresponding to 1 count per second,\\
TOT\_EXPTIME, the total exposure time that each final image represents, and\\
GAIN\_TOT, the number of photoelectrons represented by each count in 
the final image.

\section{PHOTOMETRY}
\label{sec:photometry}
 
\subsection{Photometric Calibration}
\label{sec:photcal}

Our calibration scheme is to take $\sim$5 minute exposures of our fields 
(referred to as ``calibration images'') 
in all optical filters on photometric nights.  This was achieved for 
E-HDFS on Oct. 6, 2002.  
Photometric calibration is performed 
using 
a
version of the Landolt catalog 
with all magnitudes and colors converted into the AB95 system of 
\citet{fukugitaetal96} (referred to throughout this paper 
as AB magnitudes).
We included $z'$ magnitudes for Landolt standard stars in our 
catalog of standards using the formula
\begin{equation}
z'_{\mathrm AB} = I_{\mathrm AB} - 0.59 (R-I)_{\mathrm AB} 
+ 0.16 (V-R)_{\mathrm AB} - 0.04 \; \; \; ,
\label{eq:z}
\end{equation}
which was generated from the formulae of \citet{fukugitaetal96} 
and has been corrected for a 0.02 magnitude bias found when 
comparing this prediction with measurements of $z'$ magnitudes 
of bright Landolt stars by \citet{smithetal02}.  
The resulting 0.04 magnitude rms error was used to predict 
errors in these estimates.  
This procedure offers a 
significant reduction in observing time and a tremendous 
increase in the number of calibration stars versus the traditional 
method of using spectrophotometric standard stars for filters outside 
the Johnson-Cousins system.  
Only a 
few stars per Landolt field have been turned into SDSS calibrators 
\citep{smithetal02} 
and they are typically too bright for the 4m telescope to use 
without defocusing.  

Table \ref{tab:photcal} lists the airmass and seeing for  
each of our calibration images along with our photometric 
solution for the magntiude in terms of measured counts per second, 
\begin{equation}
m = -2.5 \log_{10}(counts/s) + Z - c X - Color term \; \; \; ,
\label{eq:photcal}
\end{equation}
where $Z$ is the magnitude zeropoint, $c$ is the airmass coefficient, 
$X$ is airmass, and the color term is listed in Table \ref{tab:photcal}.  
  Our images of Landolt standard fields occupied a range of airmasses 
bracketing the airmasses of these calibration images but insufficient 
to break the degeneracy between zeropoint and airmass coefficient.  We 
therefore used fixed airmass coefficients, with the resulting uncertainty 
reflected in the uncertainty in the zeropoint fit, which is $<$1.5\% in 
each filter.  
Since AB95 was carefully calibrated for the standard Johnson-Cousins 
$UBVRI$ filter set, the conversion of the Landolt 
catalog is quite accurate.  
However, the filter system used at CTIO is not a precise match to 
Johnson-Cousins, making the use of a color term helpful in photometric 
calibration.  The relatively small color coefficients shown in 
Table \ref{tab:photcal} minimize the scatter in our photometric 
solution by providing a better estimate of the true AB magnitude of
each Landolt star in our observed filter system.  
We set the 
AB color coefficient to zero when determining photometry for our 
final object catalog; this places the fluxes on our observed filter 
system rather than standard Johnson-Cousins filters.
  A color correction can be calculated later 
for applications that require fluxes extrapolated to Johnson-Cousins 
colors.
Because photometric redshift codes 
multiply spectral templates by the atmospheric transmission, filter
transmission, and CCD quantum efficiency used for the observations, they 
already account for the origin of the color term and should be 
given fluxes in the observed filter system.  The presence of non-power-law 
features such as Lyman and Balmer breaks in these template spectra 
(and real objects) makes it desirable to avoid 
color transformations calibrated with stars.

We then used the calibration images 
to create standard stars in each filter for E-HDFS 
and used these new standard stars to calculate the zeropoint 
in our final images.  
This avoids attempting to explicitly correct the photometry 
for the airmass and cirrus extinction in each image 
which contributed to our final images, allowing us to use images taken 
in non-photometric conditions as part of the final stack.  
For E-HDFS we used a set of 10 stars, each from a different original 
amplifier, 
chosen to be bright enough to 
provide good signal in $U$ and yet to avoid saturation in 
$R$, $I$, and $z'$.    
Care was 
taken to choose some stars near the field center and some near the sides 
and  corners of the field to check for systematic effects.  
The photometry of the new standards was determined 
using the IRAF routine 
{\it phot}.
We used the same $14''$ diameter aperture for photometry of these 
standard stars in order to minimize aperture losses.  
We repeated this process by using {\it phot} on our final science 
images for the same stars.  
We checked for systematic patterns in the offsets but found none, and 
the rms scatter was 0.03 magnitudes, implying a smaller error in 
the mean offset.       

The total exposure time, seeing, photometric zeropoints, and 
 5 $\sigma$ point source 
detection depths (AB) of our final images are given in Table \ref{tab:final}.
Flux zeropoints are also determined using the 
definition \citep{fukugitaetal96} 
\begin{equation}
 f_\nu(\mu \mathrm{Jy}) = 3.631\times 10^9 \cdot 10^{-0.4 m_{\mathrm{AB}}}   \; \; \; .
\label{eq:flux_zeropoint}
\end{equation}
Given the measured uncertainties in the initial zeropoints and our 
measured scatter in the solution for the zeropoints of the final images, 
we estimate a total 
systematic uncertainty of 3\% in the zeropoint of each filter.

\subsection{Correlated Noise on Small and Large Scales}
\label{sec:correlated}

Derivations of optimal photometric techniques and their uncertainties 
typically assume Poisson noise which is uncorrelated between image 
pixels.  However, the astrometric reprojection used to place all images 
on a common grid correlates neighboring pixels, leaving the large-scale 
noise properties unchanged but 
making a pixel-by-pixel 
rms a poor measure of the typical noise due to sky fluctuations in 
a larger aperture.  As this is the method used by SExtractor to 
estimate photometric errors, we undertook a detailed investigation 
to test its accuracy, which turns out to be poor, and to find an 
improved method.  Our empirical method for estimating photometric 
uncertainties as a function of aperture size is also sensitive to large-scale 
noise correlations from unsubtracted nearby objects and from 
any CCD defects that survive flat-fielding.  
Wings
from bright objects appear to affect the noise statistics;  when we set
all pixels belonging to detected objects to zero and displayed the resultant
image, wings are clearly visible around the brightest objects.  

\begin{figure}[h]
\epsscale{1.0}
\plotone{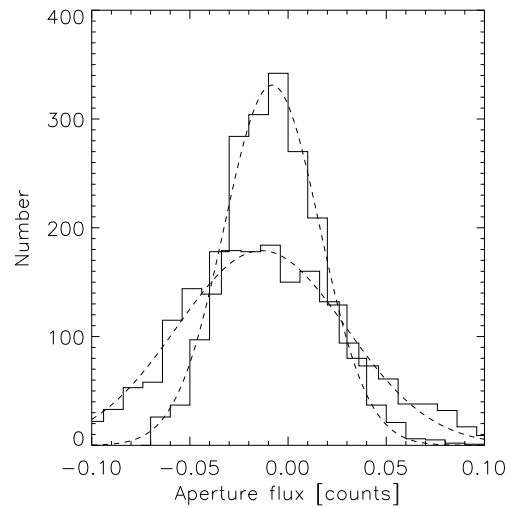}
\caption{
Histogram of aperture counts for two different aperture sizes showing 
Gaussian fits.  The larger aperture has more variation in counts and 
hence is fit by a Gaussian with larger fwhm and lower mode.  
\label{fig:hist}
}
\end{figure}

We estimate the error due to background fluctuations and 
noise correlations in each filter via a custom IDL
code which places $\sim2000$ random apertures of a given size on the
sky-subtracted image which do not overlap with any of the pixels in
the segmentation map of isophotal object regions produced by 
SExtractor.  
We use circular apertures of area $n_{pix}$ centered 
at integer pixels and describe them 
by an effective size $N = \sqrt{n_{pix}}$.   
A Gaussian is then fit to the histogram of aperture fluxes
to yield the rms background fluctuation as shown in 
Fig. \ref{fig:hist}.  
The histograms appear well
described by the best-fit Gaussians.

\begin{figure}[ht]
\epsscale{1.0}
\plotone{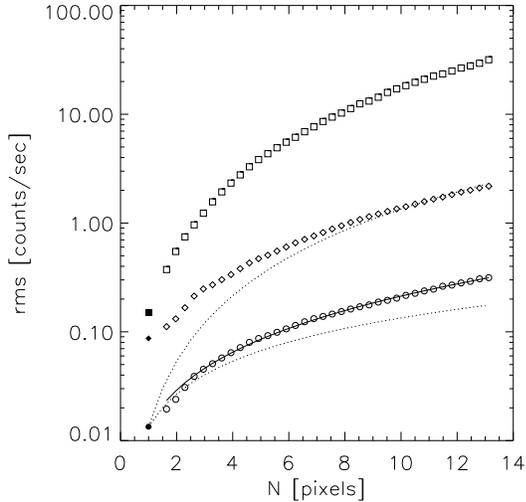}
\caption{
V-band rms fluctuation in counts/s as a function of linear aperture 
size $N$ in pixels for a raw image (squares), sky-subtracted 
raw image (diamonds) and the final image (circles).  Filled symbols 
show $\sigma_1$ for each image which is calculated pixel-by-pixel rather 
than using a circular aperture.  For the final image, the solid 
line gives the fit of Eq. \ref{eq:power} and the dotted lines show 
the limiting cases of rms proportional to $N^2$ and $N$.     
\label{fig:sigmaV}
}
\end{figure}

Figure \ref{fig:sigmaV} shows the rms of aperture fluxes  
vs. N for raw, sky-subtracted raw, and final 
$V$ images.  The solid curve gives our fit to the noise 
properties of the final $V$ image  using the function 
\begin{equation}
\sigma_N = \sigma_1 \alpha N^\beta \; \; \; , 
\label{eq:power}
\end{equation}
where 
$\alpha=0.77$, $\beta=1.30$ for the measured rms pixel noise of 
$\sigma_1$=0.014.  
An alternative formula suggested by \citet{labbeetal03}, 
\begin{equation}
\sigma_N = \sigma_1 a N + \sigma_1 b N^2 \; \; \; ,
\label{eq:labbe}
\end{equation}
yields an equally good fit, with 
$a=1.007$, $b=0.051$.
This formula shows an explicit sum between contributions from Poisson 
noise, which is independent from one pixel to another and hence has 
rms proportional to $N$, and correlated noise from fluctuations 
in background level on scales larger than the aperture which yields 
an rms proportional to $N^2$. 
These simple, extreme cases are shown as dotted lines starting 
from the value of $\sigma _1$ in 
Figure \ref{fig:sigmaV} and are seen to bracket the true behavior   
except at very small aperture sizes.  Apertures of just a few 
pixels in area are affected by the small-scale noise correlations   
introduced by re-pixelization performed during 
re-projection; 
this effect should actually depress $\sigma _1$ since each final 
pixel is the average over several nominally independent input pixels.  
 We prefer the simplicity of 
Eq. \ref{eq:power} which reflects the reality that noise exists on 
a range of scales leading to an effective power-law behavior intermediate 
between these two extremes.   
Table \ref{tab:noise} lists our measured rms-per-pixel $\sigma_1$   
and the fit coefficients $\alpha$ and $\beta$ for all of our filters; 
the uncertainties in $\alpha$ and $\beta$ are highly correlated but 
both have been determined to 5\% precision, whereas the 
uncertainties in $\sigma_1$ measurements are at the 1\% level.

A similar analysis of our raw images shown in 
Figure \ref{fig:sigmaV} reveals that the noise in the raw images 
(squares) is 
fully correlated due to the domination of background fluctuations.  
Once a sky-subtraction has been performed (diamonds), 
the noise properties resemble 
those of the final image except for being noisier by a factor 
roughly equal to the square root of the number of exposures.  
While some of this improvement comes from flat-fielding, it appears 
that the correlated noise is due to background fluctuations from 
objects whose subtraction improves with the square root of 
observing time.  This is inconsistent with confusion noise from 
undetected sources, which is not expected to be a 
significant factor even in optical images this deep.  
One possibility for the origin of this correlated noise 
is that the global sky subtraction performed by SExtractor is 
insufficient.  However, when we mimic the BACKPHOTO\_TYPE LOCAL mode of 
SExtractor by estimating the local sky in a 
square annulus 
around each aperture and subtracting this off we find a slight 
increase in the noise correlations.  Hence it appears that 
the background contributions from sky gradients and nearby objects 
have been subtracted about as well as one could expect, and that the 
remaining contributions increase the noise significantly above 
the extrapolation from $\sigma_1$ one would assume for the 
case of uncorrelated noise.  

\subsection{Optimal Apertures for Photometry}
\label{sec:optimal_apertures}

\begin{figure*}[t]
\plottwo{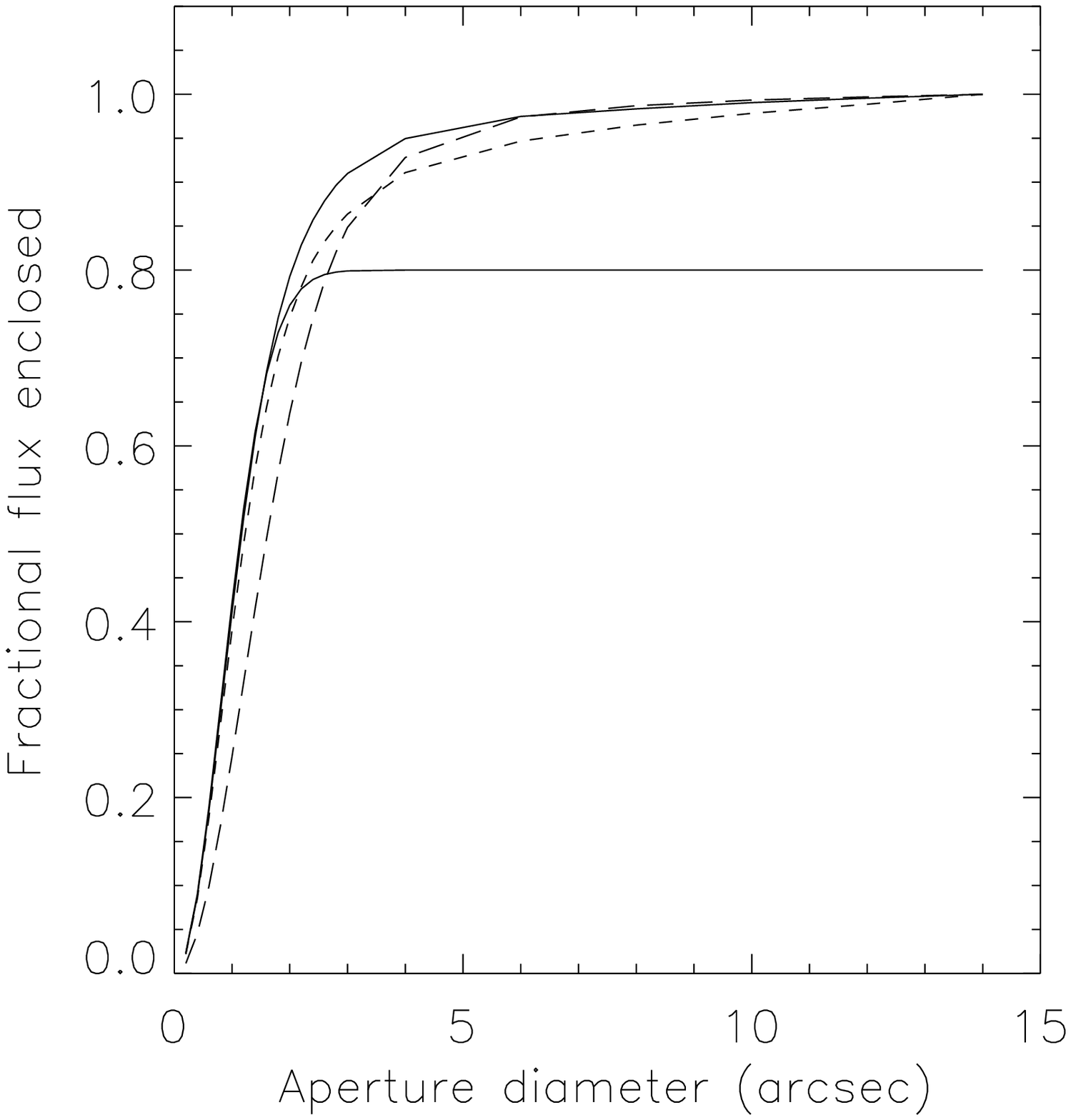}{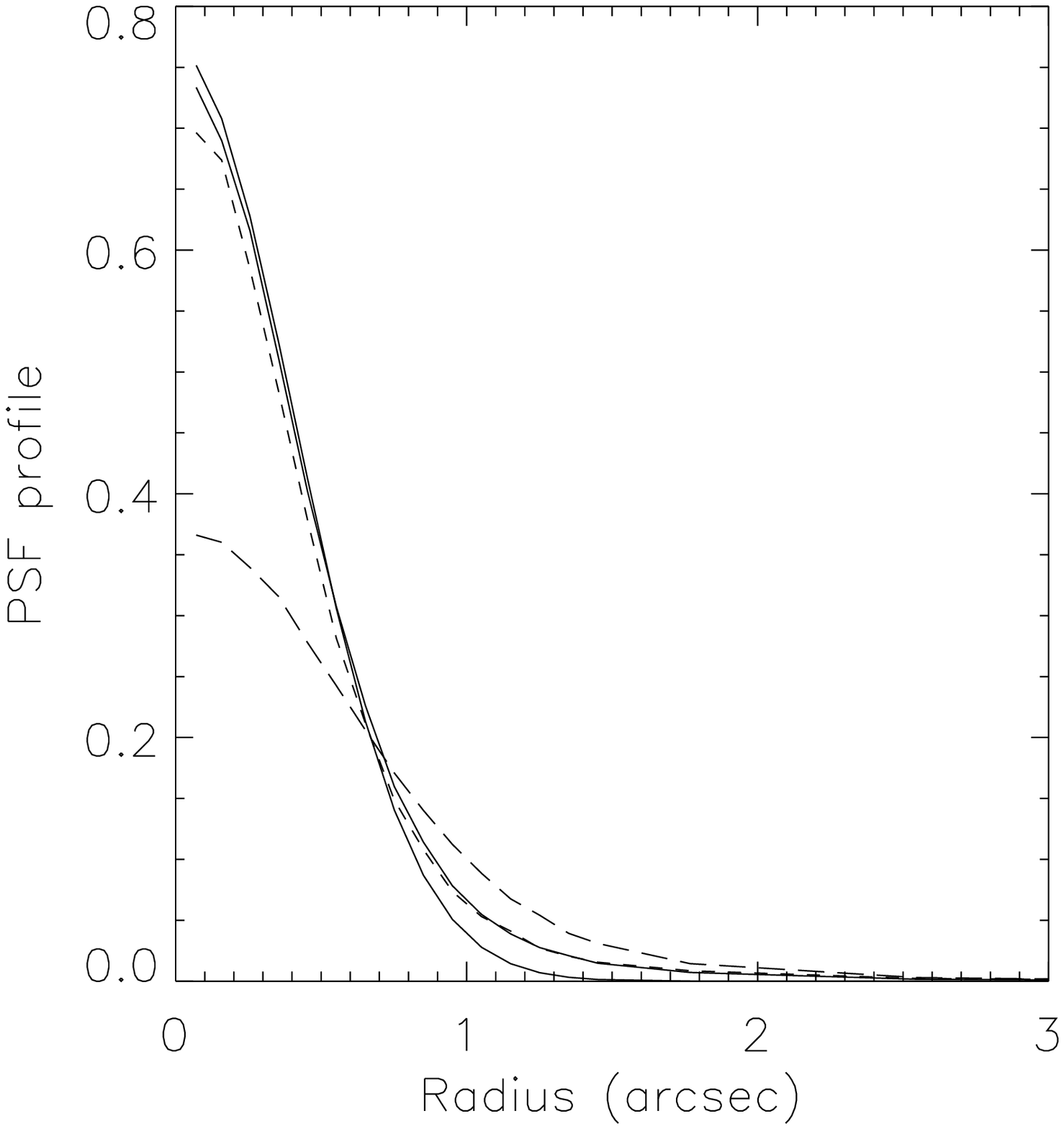} 
\caption{Left panel:  
Median fractional flux enclosed for bright, unsaturated point sources 
as a function of aperture diameter.  Right panel:  PSF implied by 
this enclosed flux profile.  In both panels, the upper solid curve shows 
results for the $V$ band, with the lower solid curve representing 
a Gaussian PSF which matches the median  fwhm of 0.96$''$ measured 
by SExtractor on the $V$ image but only contains 80\% of the total 
object flux.  This shows that the PSF core is Gaussian but the wings 
are broader.  The long-dashed curve shows results for the $U$ band, 
which has worse seeing but equally rapid convergence to the 
total object flux in the left panel.  
The short-dashed curve shows results for 
the $I$ band, which has nearly identical seeing to $V$ but the slower 
convergence to the total object flux seen in the $RIz'$ filters.
\label{fig:psf}
}
\end{figure*}

We wish to determine accurate photometry for a wide range 
of object sizes and brightnesses, with particular attention to 
high-redshift galaxies with half-light radii 0.2-0.3$''$ 
\citep{steideletal96b}) 
which are dim and typically unresolved 
in our $\sim 1''$ seeing 
and can therefore be approximated as 
point sources.  Indeed, \citet{smailetal95} showed that most 
galaxies at $m\simeq 25$ have half-light radii $< 0.3''$ 
regardless of redshift.  
In traditional aperture photometry, 
the flux of an object is measured by giving 
constant weight to one region of the image 
and zero weight to the rest, with  
fractional weights used for pixels which fall 
only partially in the chosen aperture.  The simplest case is  
a circular aperture centered at the object centroid.  
For the case of uncorrelated noise dominated by the sky 
background and a Gaussian PSF, the circular aperture 
with optimal signal-to-noise is easy to derive and has  
a diameter equal to 1.35 times the seeing fwhm.  
However, the actual PSF in each filter has a near-Gaussian 
core with broad wings i.e. more of the flux is found outside 
the fwhm than for a Gaussian.  
To investigate the effect of the non-Gaussian PSF on photometry we 
measured the enclosed flux as a function of radius for a set of 
objects selected to be bright but unsaturated ($19<m<20$) 
and unresolved according to 
SExtractor ($stellarity > 0.8$, as discussed below).  
Fig. \ref{fig:psf} shows the results for the $U$, $V$, and $I$ bands 
plus a Gaussian of 0.96$''$ fwhm that represents an excellent fit to the 
core of the $V$-band PSF but only encloses 80\% of the total flux.  
A 14$''$ diameter aperture was used to define the total object flux, 
with rapid convergence seen in the bluer filters but only gradually 
in $R$,$I$, and $z'$.  While this could be due in part to contributions 
from neighboring objects, we used the median statistic to define these 
profiles to avoid sensitivity to the small fraction of objects with 
bright neighbors.  The contribution of nearby objects should lead to 
a divergence with increasing aperture diameter which is not seen, so 
it appears that the PSF wings do contain a higher fraction of the total flux 
in the redder bands, with perhaps a few percent missed 
by our 14$''$ diameter aperture.

\begin{figure}[h!]
\epsscale{1.0}
\plotone{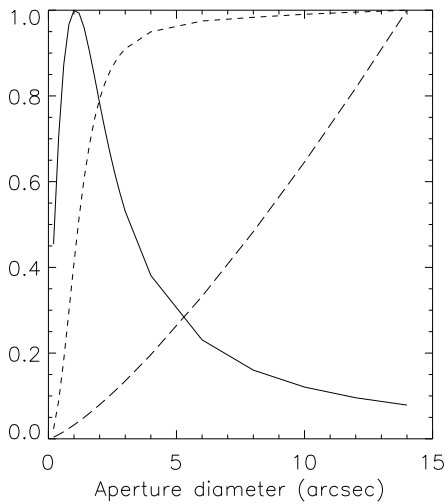}
\caption{The 
long-dashed curve shows the $V$-band noise as a function of aperture 
diameter, normalized to the value at 14$''$.  The short-dashed curve shows 
the $V$-band enclosed flux from Fig. \ref{fig:psf}.  
The solid curve shows the resulting signal-to-noise ratio, which has a
maximum at 0.5$''$ where less than half of the flux is enclosed.    
The behavior for other filters is similar. 
\label{fig:sn}
}
\end{figure}

Figure \ref{fig:sn} plots the enclosed flux, the rms background fluctuation, 
and the resulting signal-to-noise for point sources as a function of 
aperture diameter.  
There are competing effects which serve to make the optimal 
aperture slightly smaller than the idealized result of 
1.35 $\times$ fwhm.  The non-Gaussian PSF places more flux outside the 
``optimal'' diameter, so a larger aperture would find more signal.  
However, the large-scale noise correlations mean that noise 
increases with aperture size faster than in the ideal Poisson case.  
Table \ref{tab:apertures} lists various seeing measurements along with 
the optimal circular aperture diameters 
and fractional flux enclosed within these 
apertures for each filter which is used to correct the observed 
fluxes to estimate the total object flux in each filter.  
A systematic increase in the seeing 
occurs when the estimate is made using SExtractor's {\it fwhm} 
parameter, which assumes a Gaussian PSF, instead of IRAF's {\it 
imexam} routine, which allows a Moffat fit.  The non-Gaussian 
PSF shape is reflected in the further increase in the diameter 
containing half of the object light implied by the half-light radii 
$r_{1/2}$ output by SExtractor; a Gaussian contains half of the flux 
within one fwhm.  The optimal circular apertures for point sources 
turn out to be 
quite close to these values for the half-light radii.  
The image quality is measured to be nearly constant across 
the $33'\times 34.5'$ field, with systematic variations in fwhm and 
half-light 
radius seen at the 10\% level in the $I$ band but at no more 
than the 5\% level in the other filters.

For extended sources, our apertures are no longer optimal, and 
the corrections for the fraction of enclosed flux are incorrect.   
These sources are barely affected by seeing 
smaller than their intrinsic size, leading to a flux better described 
by the aperture area than by the fractional enclosed point-source flux.   
The traditional solution is to use larger apertures ($\geq 2''$ diameter) 
to measure colors of extended objects as well as point sources.  
Our analysis shows that 
the signal-to-noise of point source photometry is reduced by up to 30\%
by using these larger apertures, with a bias in the $U-V$ color of  
point sources of 20\% (0.2 magnitudes) introduced by the different seeing 
in these images.  This could be solved by convolving all of the images 
to the relatively poor seeing (fwhm=1.3$''$) of the $U$ band 
(see \citealt{labbeetal03}).  However, the non-Gaussian PSF shape requires 
a non-Gaussian convolution kernel; simply matching the fwhm by 
Gaussian convolution is not 
sufficient to fully match the PSF.    
Moreover, smoothing the other images to match the PSF shape of the 
$U$ band  would be costly in terms of decreasing 
the signal-to-noise for point source photometry, especially given 
the observed large-scale correlations in background noise.   
In choosing the ``optimal'' aperture sizes 
for each filter in Table \ref{tab:apertures}, we have chosen amongst 
the range of apertures with signal-to-noise$>$95\% of the optimal value 
in a manner that reduces the range of aperture sizes in order to reduce 
the errors introduced for extended sources.  This has the effect of 
choosing slightly larger apertures in the redder filters than would 
formally maximize the signal-to-noise.  This provides additional 
benefits, as the approximations that high-redshift galaxies are point 
sources, image co-registration is perfect, and object positions 
have no error     
all become less valid as the apertures decrease in size.  Our simulations  
indicate that these combined effects will not affect object colors beyond the 
10\% level for the chosen aperture sizes.   

Our primary solution to the challenge of obtaining decent colors for both 
unresolved and extended 
objects dimmer than the night sky 
is to create corrected aperture fluxes (hereafter referred to as 
APCORR). We tried using the  A\_IMAGE and B\_IMAGE parameters  
to model the flux distribution as an ellipsoid but encountered too 
many systematic problems with these parameters to prefer this approach.  
We therefore use the 
half-light radius (FLUX\_RADIUS) of each object determined by 
SExtractor in the $BVR$ detection image and assume a two-dimensional 
gaussian light profile with twice this value as 
fwhm.\footnote{We found FLUX\_RADIUS to be considerably 
more robust than FWHM, making it the best measure of the 
object light distributions.}  
The half-light radius measured in the $BVR$ detection image represents 
the object's intrinsic size convolved with the $BVR$ seeing, and this 
allows us to predict the observed size in each filter given the known 
seeing measured from the median half-light radii of bright point sources 
listed in Table \ref{tab:apertures}.  
For object $i$ in filter $j$ we expect  
\begin{equation}
\sigma^2_{ij} = \frac{1}{2\ln 2} 
    \left [ r^2_i + (r^2_{ps,j} - r^2_{ps,BVR}) \right ] \; \; \;,  
\label{eq:sigma_squared_ij}
\end{equation}  
where the last term in parentheses modifies the quadrature sum of 
intrinsic object size and BVR seeing to the needed sum of 
intrinsic object size and seeing in filter $j$ and the prefactor 
converts from half-light radius (which equals 
half-width half maximum for a gaussian) to the rms of the light profile.  
The fractional flux of object $i$ falling within the optimal 
circular aperture radius $r_{ap,j}$ 
for filter $j$ listed in Table \ref{tab:apertures} 
is then assumed to be 
\begin{equation}
frac_{ij} = min \left ( frac_{ps,j}, 
1 - \exp\left ( \frac{-r^2_{ap,j}}{2 \sigma^2_{ij}}\right ) \right )  
   \; \; \;, 
\label{eq:fractional_flux_ij}
\end{equation}
where the first value in parentheses refers to the fractional flux 
for point 
sources listed in Table \ref{tab:apertures} and the 
second gives the fractional flux of the two-dimensional gaussian 
which approximates the object.  Taking the minimum of 
these two values prevents 
noisy data from causing an object to be inferred to be smaller than a 
point source.  Each object's flux measured within the optimal aperture 
for a given filter is then corrected by dividing by $frac_{ij}$ to 
infer the total flux in that filter.  

Another solution for attempting to handle 
both extended 
and unresolved 
sources is SExtractor's AUTO 
aperture, which  uses a Kron ellipse adapted to each object's 
light profile to find $\simeq 94\%$ of the flux for 
extended sources and $\simeq 97\%$ for point sources using the 
parameters we have adopted,  according to \citet{bertina96}.  
In the $U$, $B$, and $V$ filters, 
we find that AUTO indeed measures 96\% of the point source flux 
found in a 14$''$ diameter aperture, with the value decreasing 
to 93\% in $R$, 92\% in $I$, and 89\% in $z'$ in keeping with the 
broader PSF discussed above.

\subsection{Estimation of Photometric Uncertainties}
\label{sec:photerr}

Because SExtractor assumes Poisson sky noise, it produces 
underestimated photometric uncertainties for objects dimmer 
than the background sky.\footnote{We found it necessary to run 
SExtractor with WEIGHT\_TYPE BACKGROUND,BACKGROUND in order to get it to use 
the background in the $BVR$ detection image as a source of uncertainty in object detection and the background in the measurement image as a source of 
uncertainty in the photometry.}  
The standard formula 
used by SExtractor to predict photometric errors in flux  
as a function of aperture size is 
\begin{equation}
\sigma^2_{phot,SE} =  \sigma_1^2 n_{pix} + \frac{F}{GAIN} \;\;\; ,
\label{eq:uncertainty_SE}
\end{equation}
where the first term represents sky fluctuations assuming uncorrelated 
noise between pixels, $F$ is the object flux in counts within 
the aperture, and $GAIN$ is the total effective gain used to turn 
counts in the normalized image into original photons detected by the CCD, 
with the fraction $F/GAIN$ giving the Poisson variance. SExtractor 
measures $\sigma_1$ using the median fluctuation in its internally 
computed background map, and this measurement agrees with ours 
to within a few percent.
We modified this to make use 
of our fits for (correlated) background noise which yields    
\begin{equation}
\sigma^2_{phot} = \sigma_1^2 \alpha^2 n_{pix}^\beta + \frac{F}{GAIN}  \;\;\; ,
\label{eq:uncertainty}
\end{equation}
where the first term now represents empirical sky fluctuations which include 
the subdominant contribution from electronic readout noise.
Both noise terms will be larger in regions of the image where reduced 
exposure time leads to a lower effective gain and hence higher background rms.
SExtractor accounts for this by using a variance map $\sigma^2_{kl}$ which 
is either input as a WEIGHT\_IMAGE or generated internally from a 
map of background fluctuations in the image.  The errors produced by 
SExtractor as a function of position are then 
\begin{equation}
\sigma^2_{phot,SE} = \sigma^2_{kl} \left (n_{pix} + \frac{F}{\sigma^2_1 \cdot GAIN} 
\right )  \;\;\; .
\label{eq:uncertainty_full_SE}
\end{equation}
The ratio of the formulas in equations \ref{eq:uncertainty_SE} and 
\ref{eq:uncertainty} allows us to make a simple correction to the 
flux uncertainties output by SExtractor to yield
\begin{equation}
\sigma^2_{phot} = \sigma^2_{phot,SE} \left (
\frac{\alpha^2 n_{pix}^\beta + \frac{F}{\sigma^2_1 \cdot GAIN}}
{n_{pix} + \frac{F}{\sigma^2_1 \cdot GAIN}} 
\right )  \;\;\; .
\label{eq:correction}
\end{equation}
For negative fluxes, $F$ is set to zero in all of these error terms, 
since the uncertainty is entirely dominated by the background 
fluctuations.  
 For AUTO 
apertures, the area is given by that of the Kron ellipse used i.e. 
$n_{pix} = \pi  r_{kron}^2 \mathrm{A\_IMAGE~B\_IMAGE}$ 
(the so-called KRON\_RADIUS is really a scaling 
of the A\_IMAGE and B\_IMAGE parameters based on the second moment of 
the object light distribution).    

The resulting rms has units of counts per second and is converted to 
flux using the zeropoints in Table \ref{tab:final}.  
The uncertainty determined for optimized aperture fluxes is 
then multiplied by the same correction factor used to estimate the
total object flux i.e. the reciprocal of $frac_{ij}$ 
defined in Eq. \ref{eq:fractional_flux_ij}.  
Finally, the uncertainty in the corrected aperture flux is increased 
by a factor $(frac_{ps,j}/frac_{ij})^2$ which serves to 
amplify the uncertainties for extended sources to 
account for the uncertainty in their 
correction factors. 
This reproduces the errors found in corrected aperture fluxes 
for extended objects in our simulations described in \S \ref{sec:simulations}. 
The photometric errors derived in this manner do not include the 
3\% calibration uncertainty which is common to all sources in a 
given filter.

\begin{figure*}[th]
\plottwo{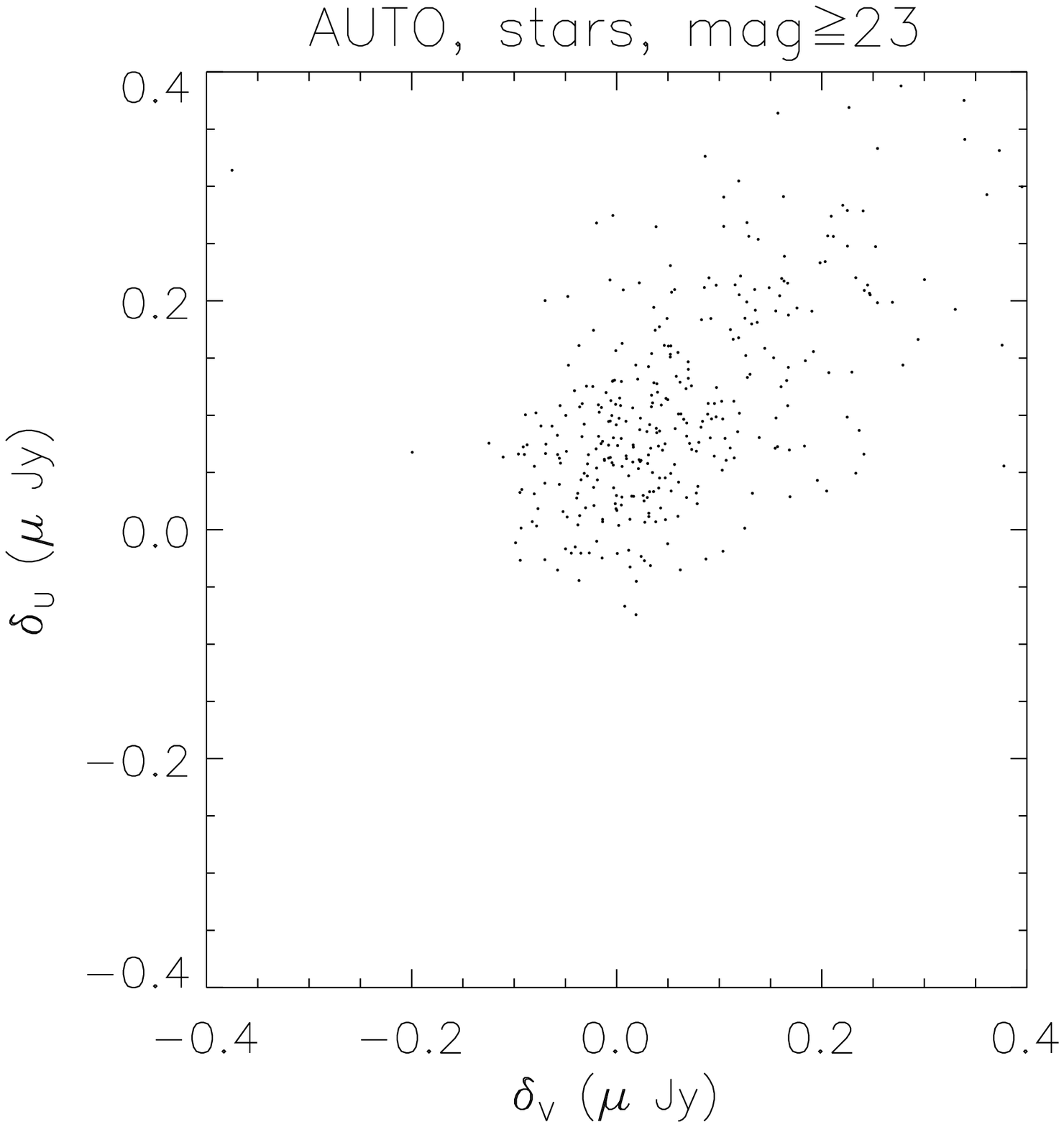}{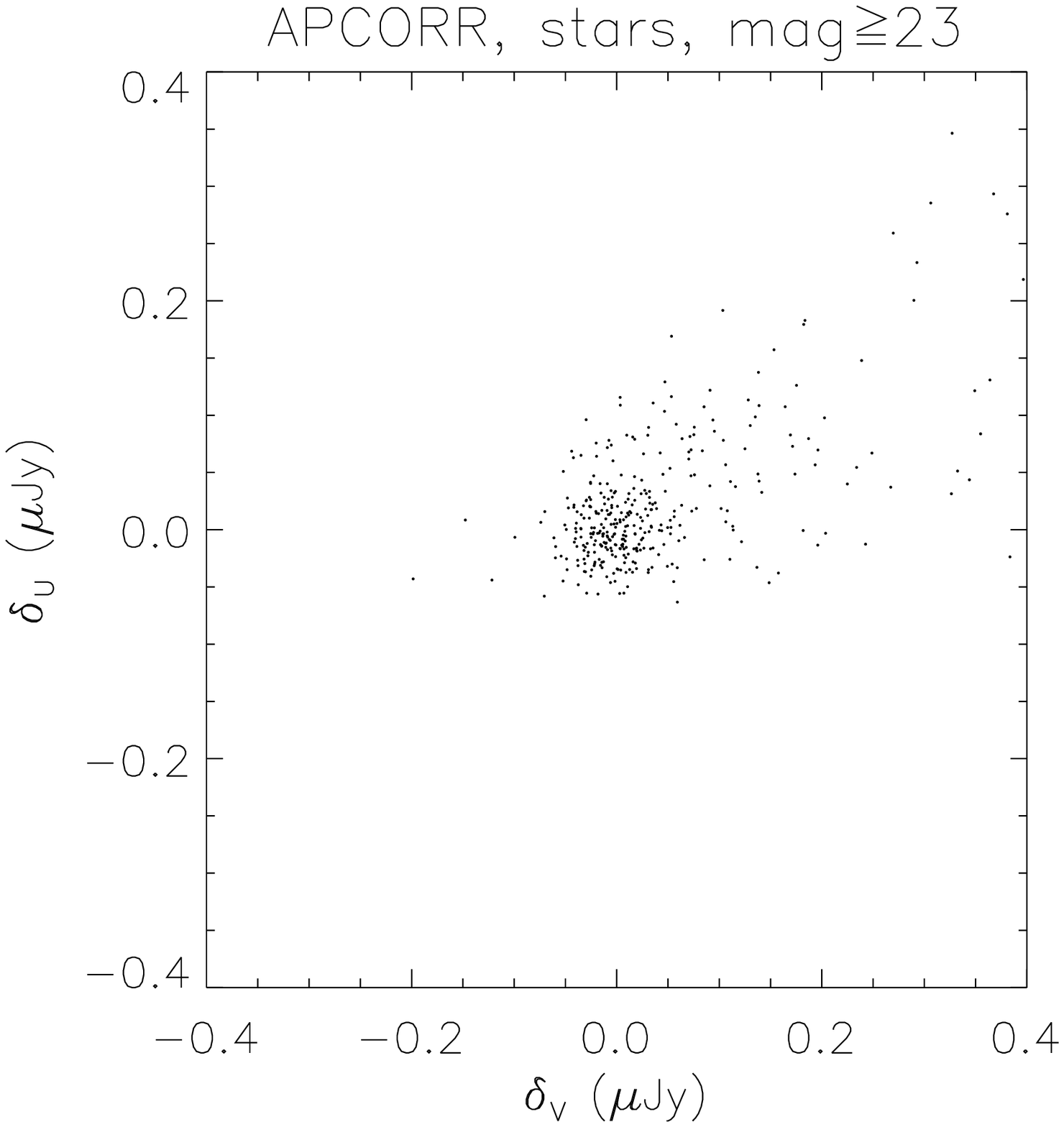}
\caption{Errors, i.e. (observed - true), 
in estimated V fluxes versus errors in estimated U fluxes for 
simulated point sources.  Left panel shows results for the 
AUTO aperture as chosen 
by SExtractor, and the right panel shows results for the 
corrected aperture fluxes described in the text.
The APCORR fluxes are unbiased 
and have a significantly smaller scatter than AUTO.
\label{fig:stars}
}
\end{figure*}

\begin{figure*}[ht]
\plottwo{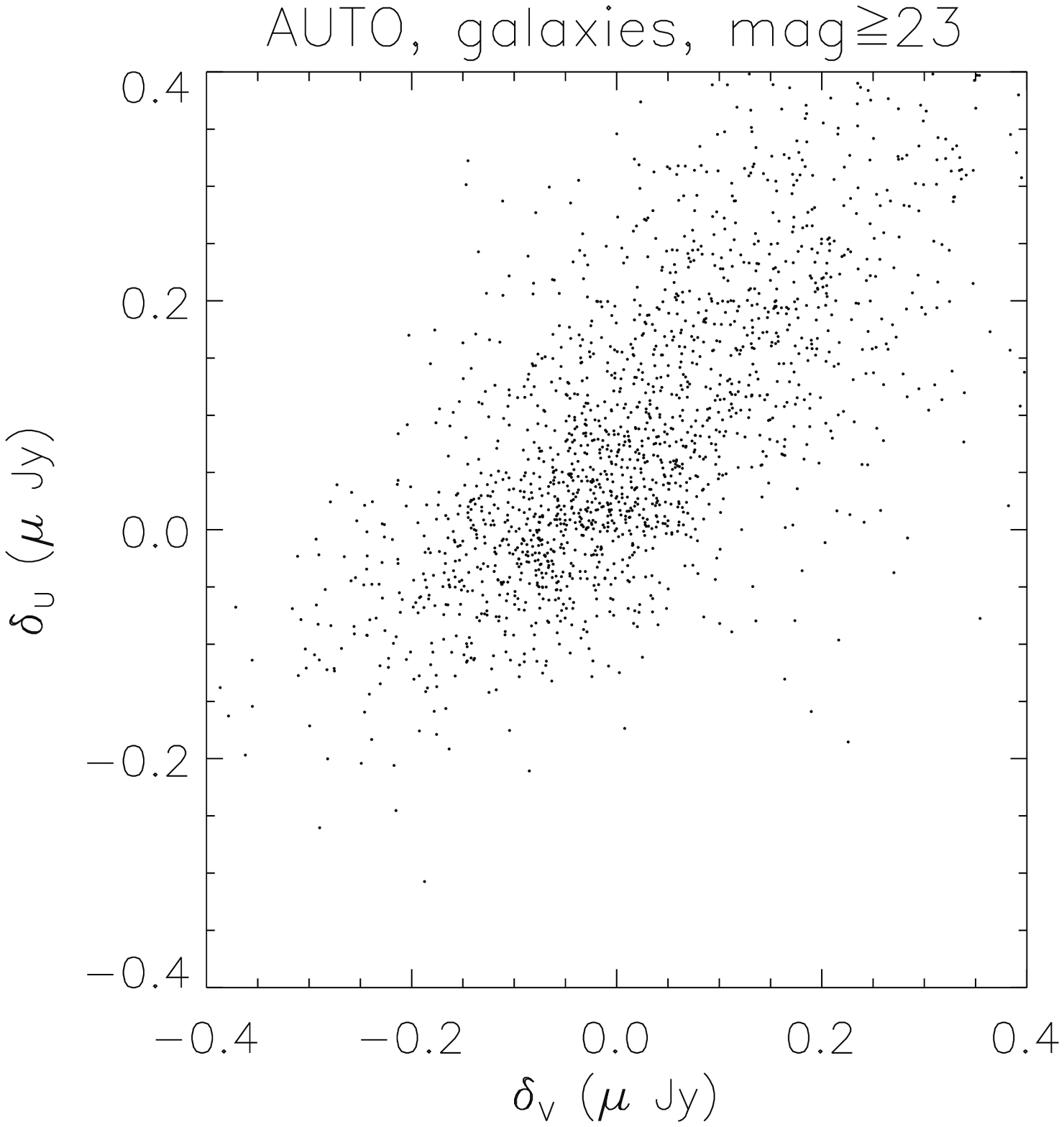}{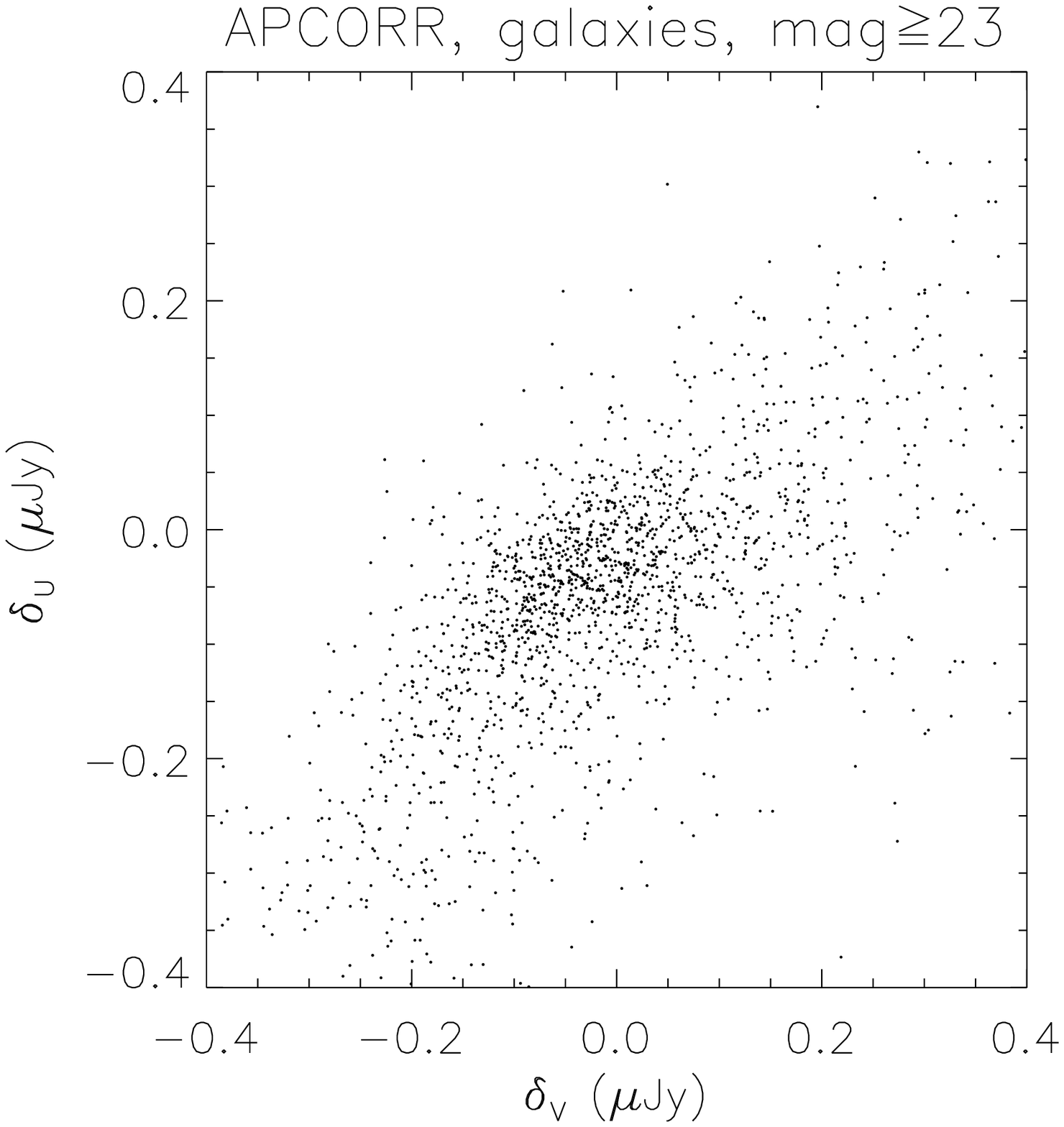}
\caption{Errors in estimated V fluxes versus errors in estimated U fluxes for 
simulated galaxies.  The left panel shows results for the 
AUTO aperture as chosen 
by SExtractor.  The right panel shows results for the 
corrected aperture fluxes described in the text.
APCORR appears nearly unbiased, and the size of the errors for galaxies are
comparable to those of AUTO but exhibit different systematics.  
\label{fig:gals}
}
\end{figure*}

\subsection{Photometry Tests on Simulated Sources} 
\label{sec:simulations}

We used the IRAF package {\it artdata} to simulate stars and 
galaxies with known magnitudes and positions and to add them 
to our observed images to get realistic crowding effects and 
background noise.\footnote{The number of simulated sources 
was roughly one-tenth that of real sources so the crowding characteristics 
of the images can be considered unchanged.}    
We ran SExtractor on these ``simulated'' images in dual-image mode to 
simulate our full photometric pipeline.  We found 
object centroiding errors to be $\leq 0.3$ pixels i.e. $\leq 0.1''$ 
at $R\leq 25$.  
in the centroiding of individual objects 
We found 
AUTO photometry to be 
nearly 
unbiased for both single filter fluxes and colors.  
However, Figure \ref{fig:stars} shows that 
AUTO has larger errors for point sources than the corrected  
aperture fluxes due to 
the larger AUTO apertures 
including significantly 
more sky noise, so we recommend AUTO fluxes 
only for extended sources and 
give fluxes for both types of apertures 
in our catalog. The errors seen in the $U$ and $V$ filters appear 
to be entirely uncorrelated.   
Figure \ref{fig:gals} shows that AUTO performs similarly to  
corrected aperture fluxes for extended 
objects, although both show significant covariance between $U$ and 
$V$ due to misestimation of the true object light profile.

\begin{figure*}[th]
\plottwo{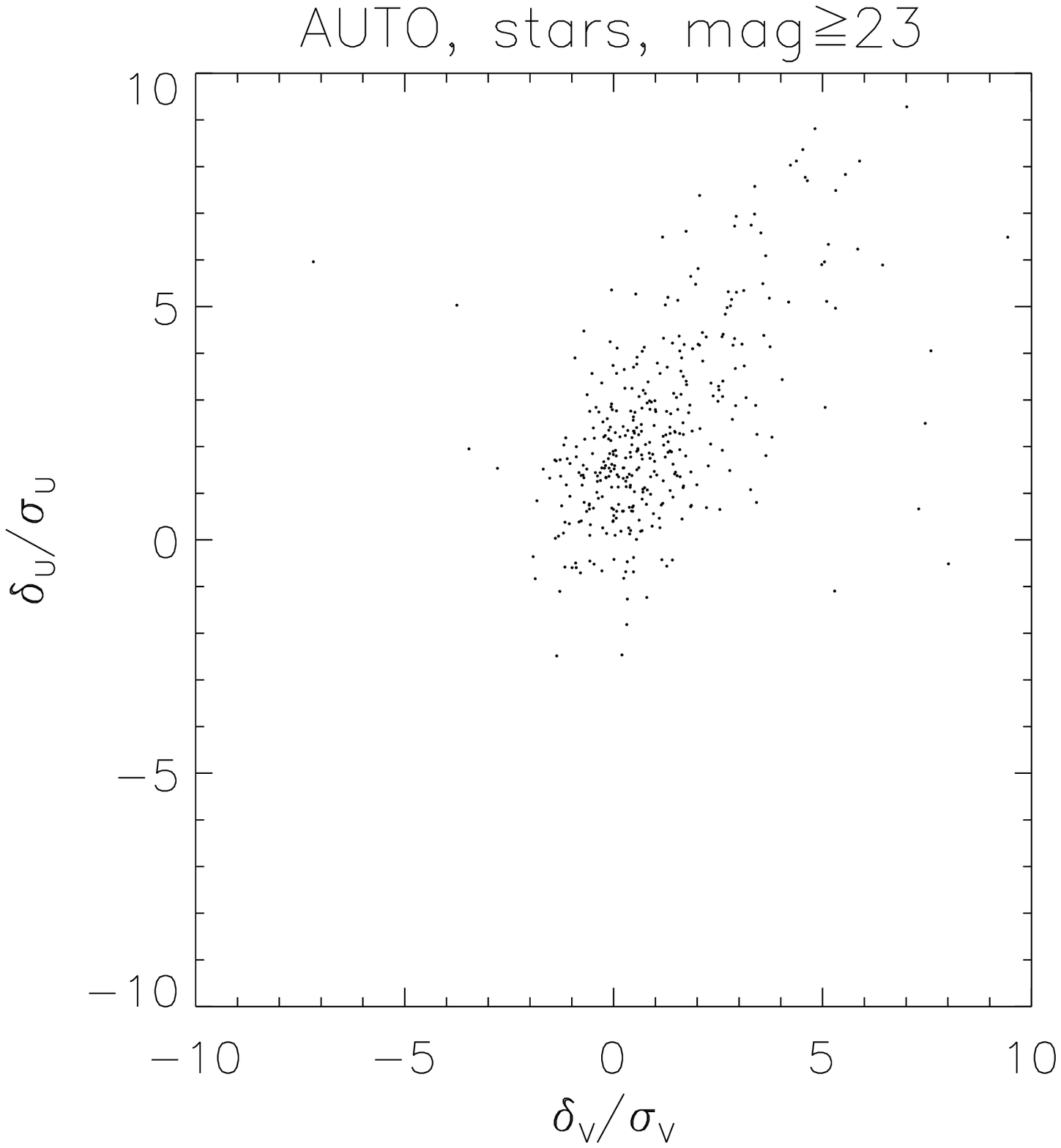}{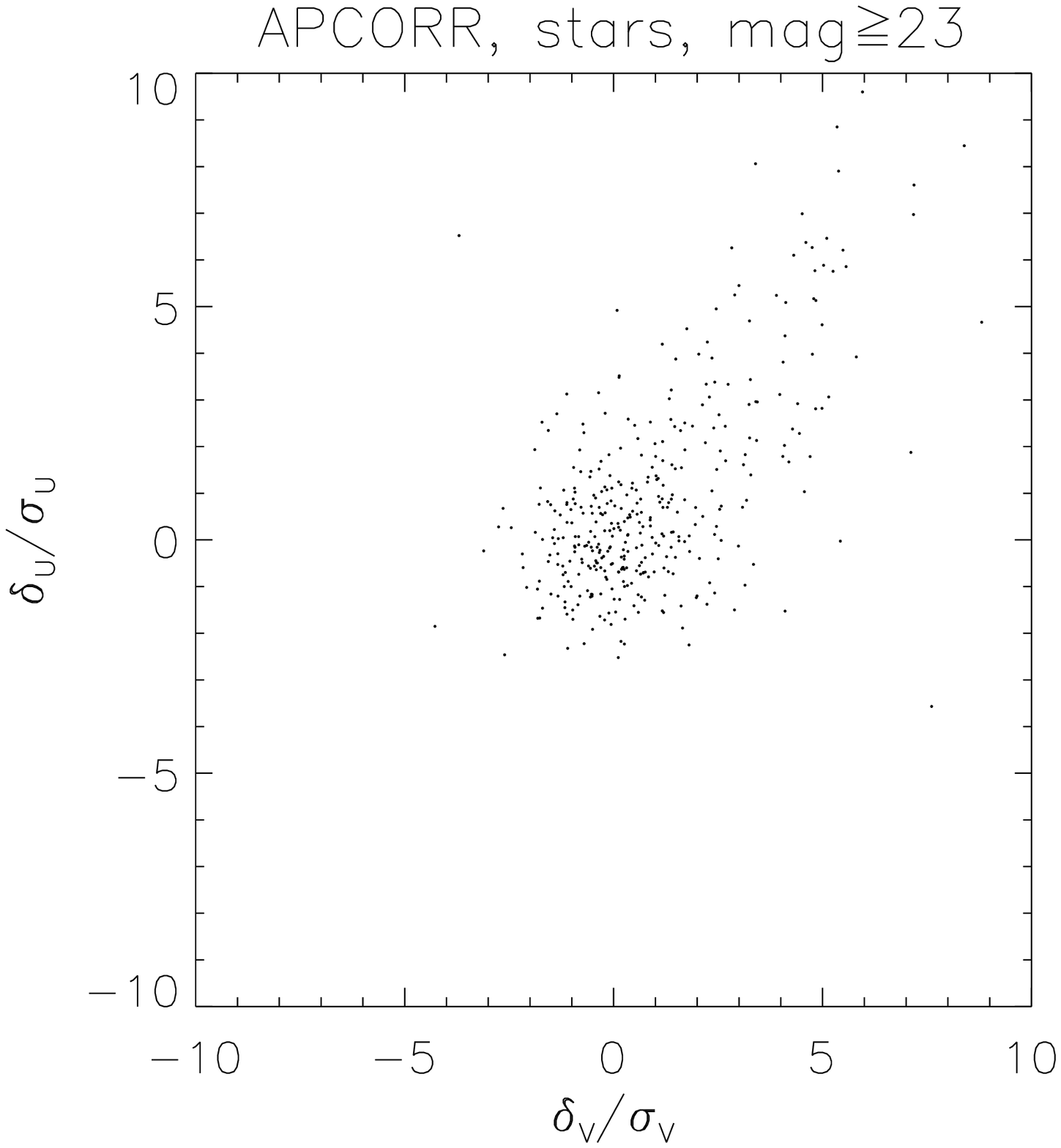}
\caption{Errors, i.e. (observed - true), 
in estimated V fluxes divided by reported $1 \sigma$ flux 
uncertainties in V vs. errors in estimated U fluxes divided by 
reported $1 \sigma$ flux uncertainties in U for 
simulated point sources.  Left panel shows results for the 
AUTO aperture as chosen 
by SExtractor, and the right panel shows results for the 
APCORR fluxes described in the text.
\label{fig:sigma_stars}
}
\end{figure*}

\begin{figure*}[ht]
\plottwo{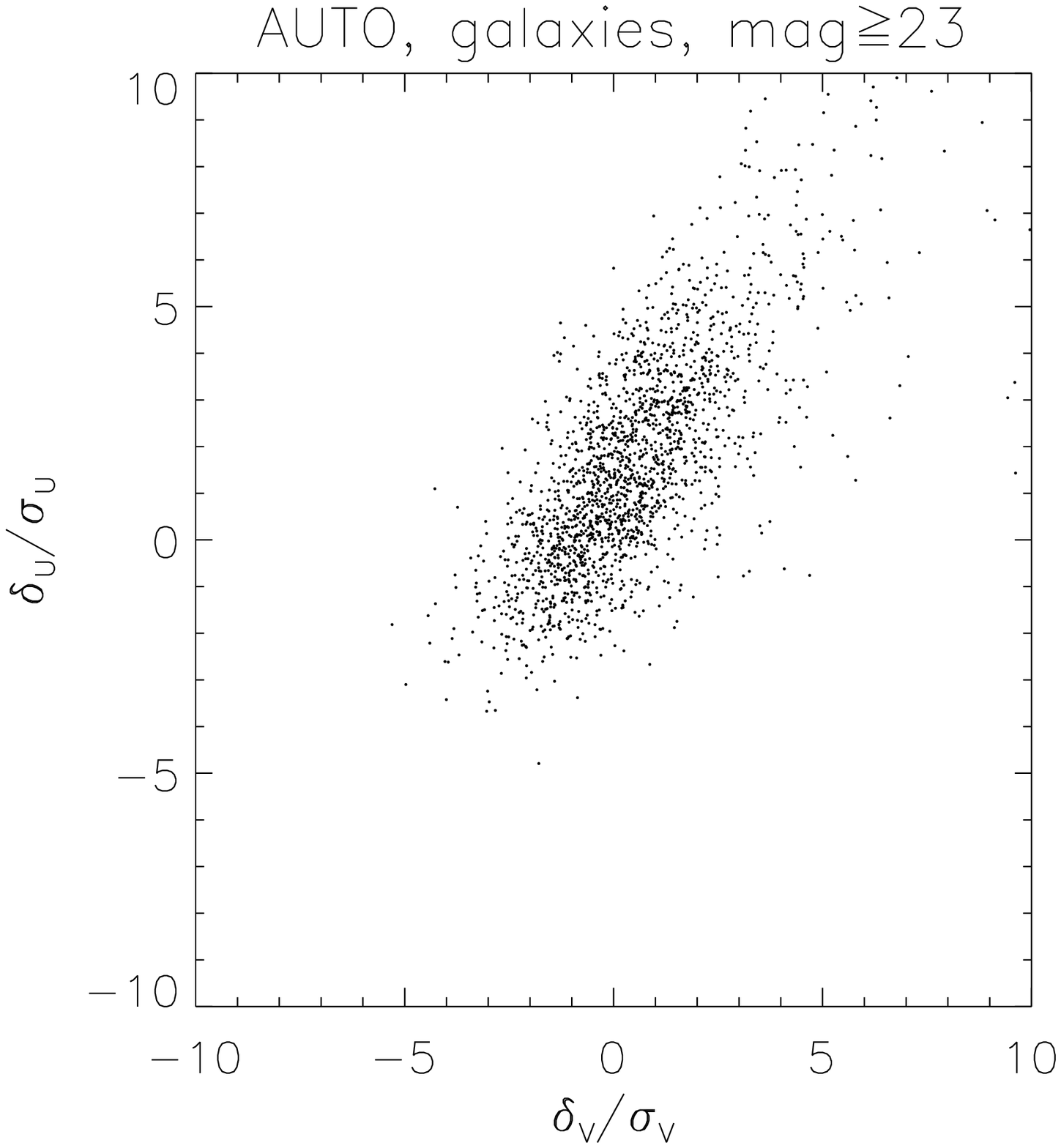}{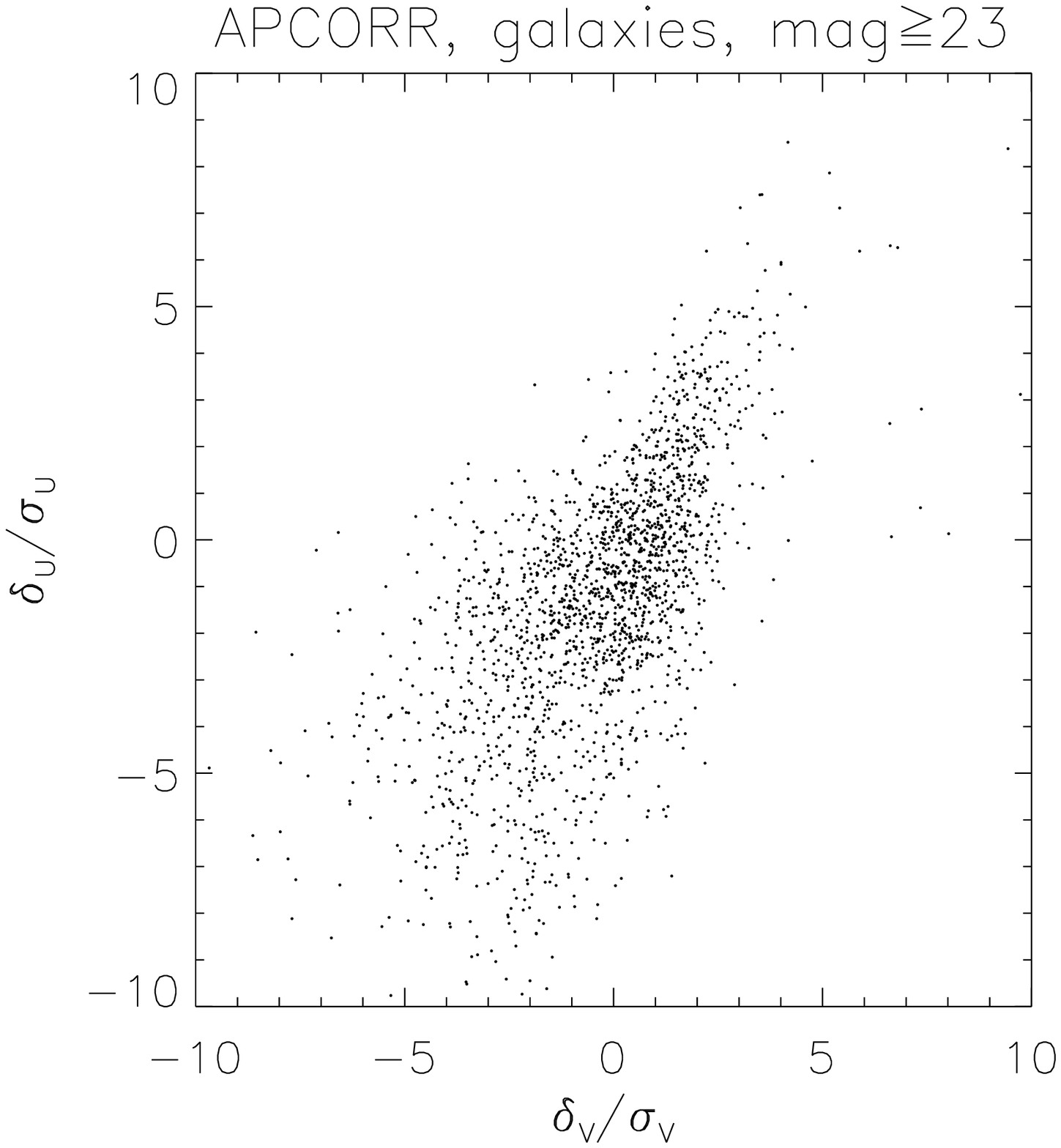}
\caption{Errors in estimated V fluxes divided by reported $1 \sigma$ flux 
uncertainties in V vs. errors in estimated U fluxes divided by 
reported $1 \sigma$ flux uncertainties in U for 
simulated galaxies.  Left panel shows results for the 
AUTO aperture as chosen 
by SExtractor, and the right panel shows results for the 
APCORR fluxes described in the text.
\label{fig:sigma_gal}
}
\end{figure*}

Figure \ref{fig:sigma_stars} shows the flux errors divided by 
their uncertainties for simulated point sources. 
APCORR again appears unbiased but this plot reveals that the 
error estimates are roughly equally accurate for both cases.  
The median squared value is close to one in both cases, so the 
reported flux uncertainties in the catalog appear trustworthy 
for point sources.
Figure \ref{fig:sigma_gal} shows the flux errors divided by 
their nominal uncertainties for simulated galaxies.  
This plot illustrates that the estimated uncertainties are 
typically too small for galaxies in both cases.

\begin{figure*}[th]
\plottwo{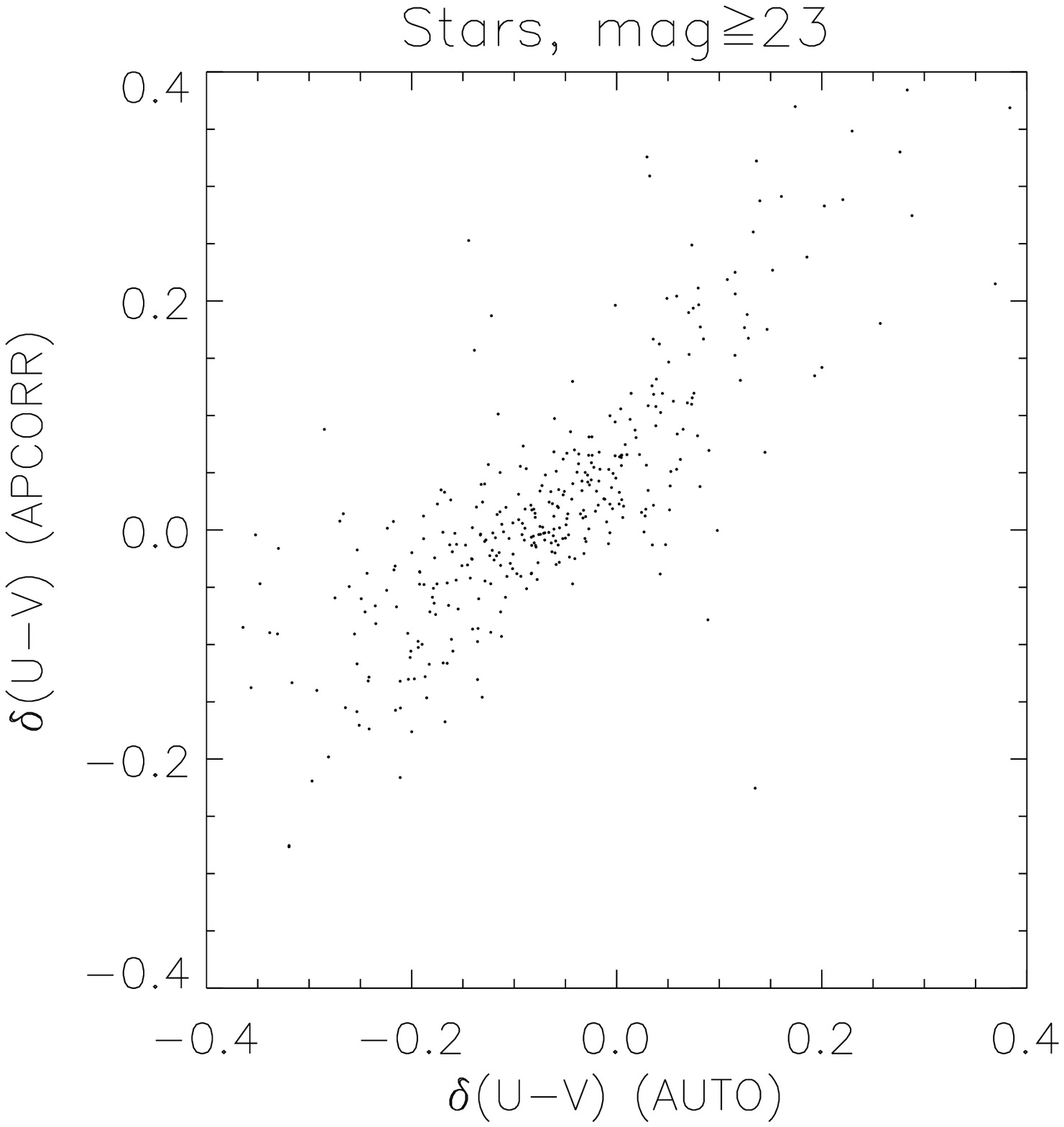}{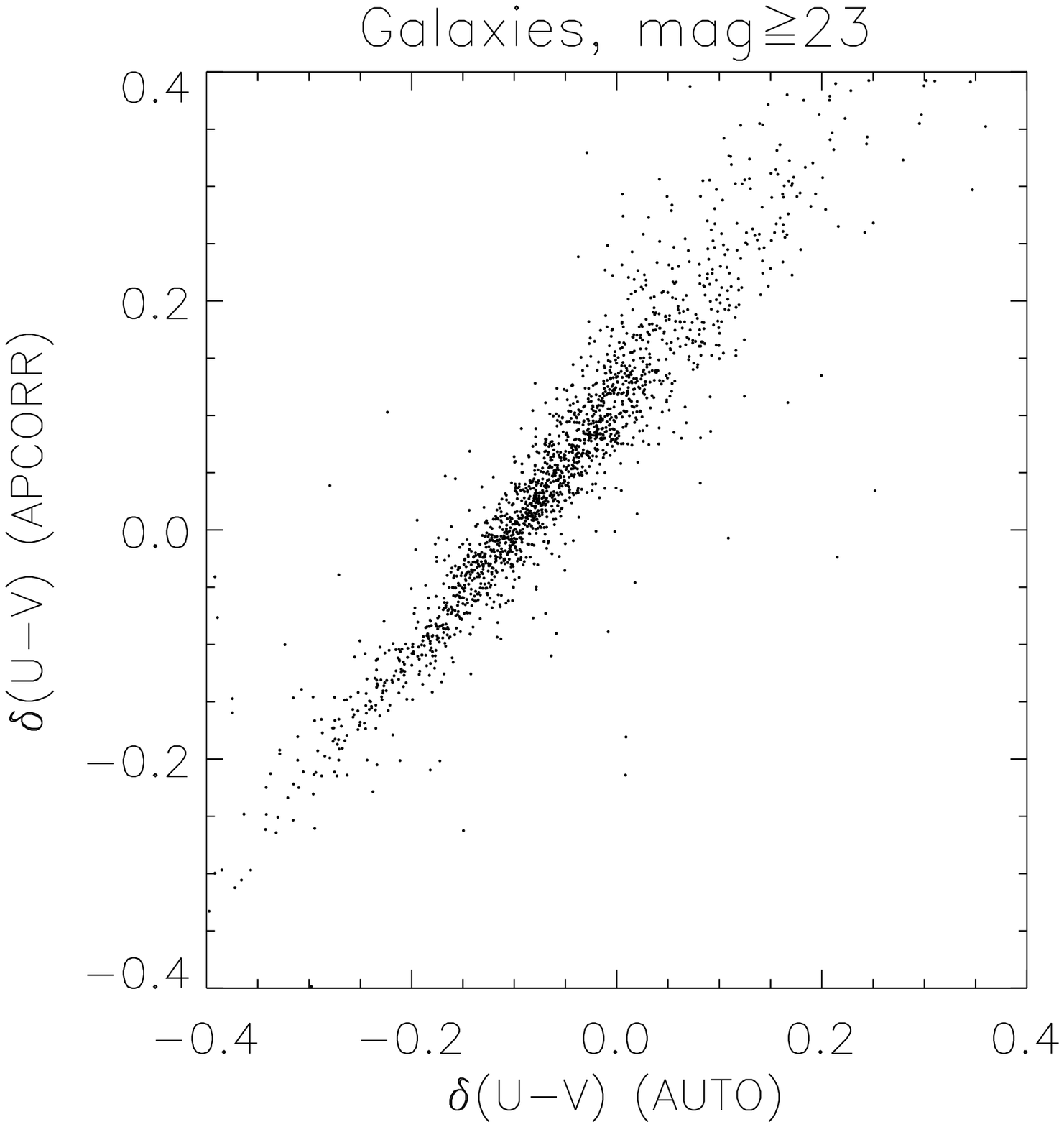}
\caption{Errors in U-V colors (AB magnitudes) for 
AUTO aperture as chosen 
by SExtractor vs. U-V color errors (AB magnitudes) for the 
APCORR fluxes described in the text.
Left panel shows results for simulated stars, and
right panel 
shows results for simulated galaxies.  
\label{fig:colors}
}
\end{figure*}

Figure \ref{fig:colors} shows
that $U-V$ colors of simulated objects 
appear slightly biased for AUTO but unbiased for APCORR, 
and the errors are a bit smaller for APCORR for point sources and 
of similar size for both types of fluxes for galaxies.  
The color errors are highly correlated, especially 
for galaxies, implying 
that these errors are primarily caused by errors in the isophotal object 
detection by SExtractor on which both fluxes depend.  These include 
problems caused by blending with nearby neighbors, although the median 
object is unaffected by neighbors in this uncrowded field.  
The AUTO and APCORR colors of real objects are discussed 
in \S \ref{sec:lbg_select}.  

\subsection{Aperture Photometry}
\label{sec:aperture}

We performed aperture photometry using SExtractor in 
dual-image mode with $BVR$ as the detection image and each final image 
as the measurement image, 
leading to a common catalog for all filters.  
Detection was performed after filtering with a $7\times7$ pixel 
gaussian convolution kernel with 4 pixel (1.07$''$) fwhm which 
is well-matched to the seeing in our $BVR$ image (see Appendix).  
We used a threshold of 1.2 for both detection and analysis; 
a single pixel at the 5 $\sigma$ level in the original 
$BVR$ image would barely make the 1.2$\sigma$ threshold after 
filtering.  Using the identical parameters, we detected 318 
objects in a negative of the $BVR$ image, so given the symmetrical 
nature of sky fluctuations and readnoise we estimate an equivalent 
number of spurious objects in our final catalog i.e. 0.5\%.    
  For each filter, 
we used the ``optimal'' apertures and corrections described 
above as well as the AUTO aperture determined for each object 
by SExtractor  
centered on the $BVR$ barycenter.  
The photometry was corrected for Galactic dust extinction of E(B-V)=0.03  
\citep{schlegelfd98}.     

The 5$\sigma$ point source detection depths in Table \ref{tab:apertures} were 
determined by multiplying the $\sigma$ of the measured sky noise 
for the optimal point source aperture sizes by 5 and then correcting 
for the flux missed by these apertures to turn 
sky fluctuations into total point source magnitudes.
Many estimates of point source detection limits in the literature are 
based on extrapolating the rms pixel noise, $\sigma_1$, to the chosen 
aperture size assuming uncorrelated sky noise.  For correlated noise 
like that in our images, which we expect is typical, this will overestimate 
the true depth in a 2$''$ diameter aperture by 0.4 magnitudes and 
in a 3$''$ diameter aperture by 0.6 magnitudes.   

\subsection{The Photometric Catalog}
\label{sec:catalog}

Our optical photometric catalog of 62968 objects in E-HDFS 
is available in full in the electronic 
version of the journal and from our website, 
with the first several lines shown in 
Table \ref{tab:catalog}.  
All positions and shape parameters are measured from the final $BVR$ image.  
The columns in the catalog offer the following information:\\
{\it Column 1}:  Object number starting with first object as 0\\
{\it Column 2}:  Object name (MUSYC-)\\
{\it Column 3}:  Right ascension in decimal hours, double precision \\
{\it Column 4}:  Declination in decimal degrees, double precision \\
{\it Column 5}:  x barycenter \\
{\it Column 6}:  y barycenter \\
{\it Column 7}:  stellarity classification \\
{\it Column 8}:  half-light radius \\
{\it Column 9}:  rms of flux distribution along major axis\\
{\it Column 10}:  rms of flux distribution along minor axis \\
{\it Column 11}:  object position angle counterclockwise from x-axis \\
{\it Column 12}:  flags output by SExtractor\\
{\it Column 13,15,17,19,21,23}:  AUTO flux in $U$,$B$,$V$,$R$,$I$,$z'$ \\
{\it Column 14,16,18,20,22,24}:  AUTO flux uncertainty in 
 $U$,$B$,$V$,$R$,$I$,$z'$ from (\ref{eq:correction})\\
{\it Column 25,27,29,31,33,35}:  APCORR flux in $U$,$B$,$V$,$R$,$I$,$z'$ 
after division by correction factor 
based on half-light radius described in Eq. \ref{eq:fractional_flux_ij}\\
{\it Column 26,28,30,32,34,36}:  APCORR flux uncertainty in 
 $U$,$B$,$V$,$R$,$I$,$z'$ from (\ref{eq:correction}) after division by 
correction factors based on half-light radius described in 
Eq. \ref{eq:fractional_flux_ij} and multiplication by additional factor 
for extended sources described above\\

Photometric measures in the catalog are given in units of flux density in 
$\mu$Jy (1 $\mu$Jy = $10^{-32}$ W m$^{-2}$ Hz$^{-1}$). 
 The conversion to AB magnitudes is simple using the formula
\begin{equation}
AB = 23.90 - 2.5 \log_{10} f_{\nu} 
\end{equation}
\citep{fukugitaetal96}.  
  This avoids 
the loss of information that comes from SExtractor representing objects 
with negative aperture fluxes as having $m=99$ and the confusion that 
results from flux errors consistent with a non-detection being turned into 
extremely large magnitudes.  Photometric errors are nearly symmetrical 
in flux but not in magnitudes, even for objects with low 
signal-to-noise, making it acceptable to represent the uncertainties 
with a single number.  
This makes flux density the natural unit to be used for calculating 
photometric redshifts if one wishes to use a $\chi^2$ function 
to compute the likelihood as is done by the publicly available 
codes.   
For color selection, one can either turn color criteria
 into desired flux 
ratios between filters or turn the measured fluxes into 
magnitudes and subtract to obtain colors.

\begin{figure*}[t]
\plottwo{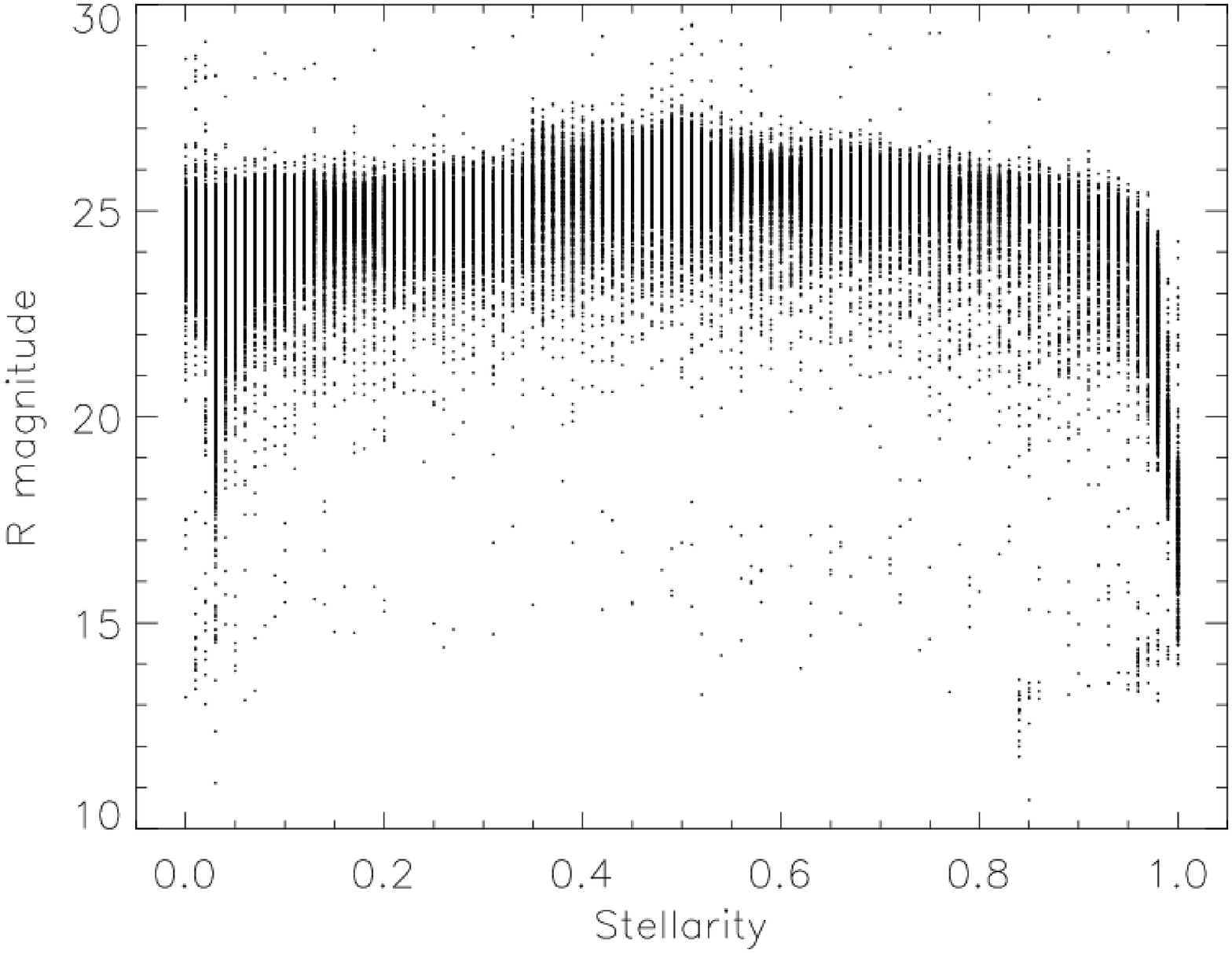}{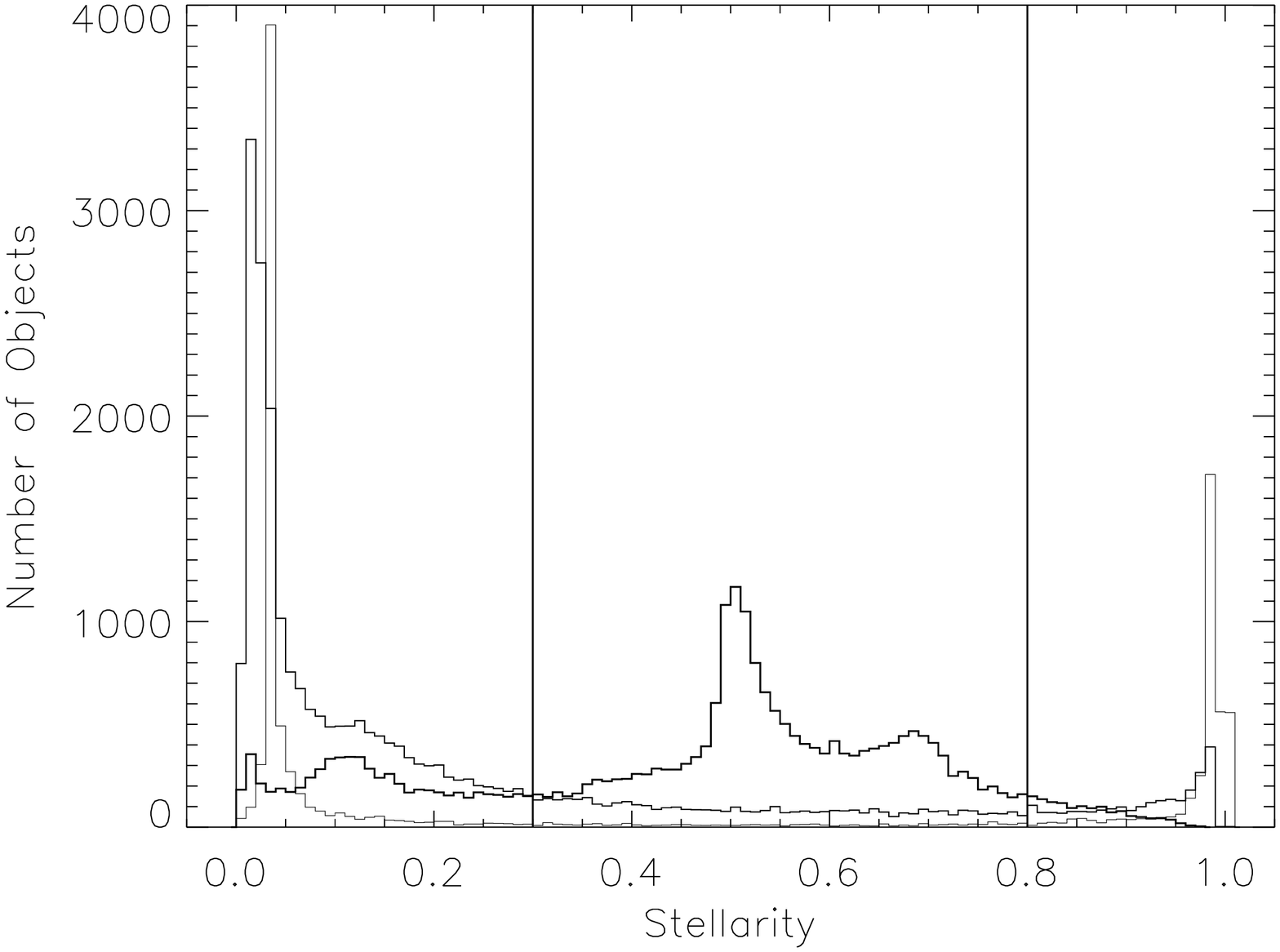}
\caption{SExtractor stellarity classification for objects in our 
catalog.  The left panel shows stellarity versus total R-band AB magnitude 
and reflects confusion for saturated objects at $R<19$. 
The right 
panel shows a histogram of stellarity, with vertical solid lines 
separating 
galaxies ($stellarity \leq$0.3), 
uncertain objects (0.3$<stellarity<$0.8), 
and
 stars ($stellarity \geq$0.8).   
The classification is clean 
at $R<23$ 
(light histogram), shows an  increasing fraction of uncertain objects 
 at $23\leq R<25$ (medium histogram), and breaks down 
at $R\geq 25$ (dark histogram).  
\label{fig:class}
}
\end{figure*}

SExtractor offers a neural network classification of objects 
into ``stars'' and galaxies; $stellarity > 0.8$ implies an unresolved 
profile consistent with the PSF, $stellarity < 0.3$ implies an extended 
profile, and $0.3 < stellarity < 0.8$ represents uncertain objects which 
are typically too dim for this classification to work successfully.  
Figure \ref{fig:class} shows the stellarity results 
for our catalog.  
The classification starts to break down at $m \geq 23$, 
with brighter objects confidently split between stellar and extended 
profiles but many dimmer objects receiving an inconclusive value.  
Stellarity becomes nearly useless at $m \geq 25$ where the majority 
of values fall in the uncertain regime, which 
is appropriate given that $r_{1/2}<0.3''$ for 
most galaxies at $m>25$ \citep{smailetal95}.  
However, the vast majority 
of objects at $m \geq 23$ are galaxies due to the rapidly rising 
galaxy luminosity function and the plateau in the stellar equivalent.  
This means that for many scientific goals all objects with 
$stellarity < 0.8$ may be considered galaxies.  

\begin{figure}[t!]
\epsscale{1.25}
\plotone{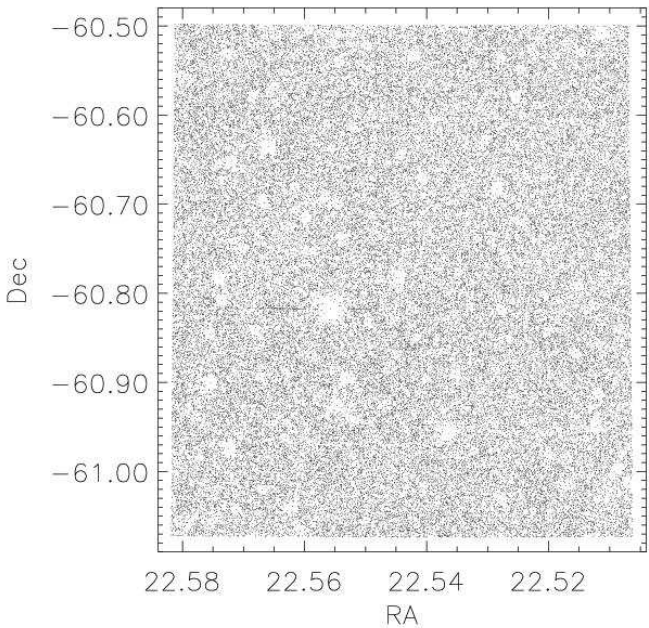}
\caption{
Sky distribution of the 62968 objects in our photometric catalog (points) 
showing regions effectively masked out due to bright stars.
\label{fig:radec}
}
\end{figure}

 Figure \ref{fig:radec} shows the locations of all objects in our 
catalog.  
There is an overabundance of objects along the image border which is 
not apparent to the eye but can be eliminated by 
removing the 574 objects with $flags \geq 8$, which 
identifies objects truncated by the image border.  
The one visual blemish is a line of spurious objects detected 
along a heavily saturated column extending from the brightest star in 
the field.  This line is removed by eliminating all 1226 objects with 
$flags \geq 4$, which implies saturation in at least one 
filter.\footnote{The 
flags column in the catalog is the result of a maximum performed 
over the flag values output by SExtractor for each measurement image run in 
dual-image mode.}

\section{RESULTS}
\label{sec:results}

\subsection{$R$-band number counts}
\label{sec:counts}

\begin{figure}[b!]
\includegraphics[angle=270,scale=0.35]{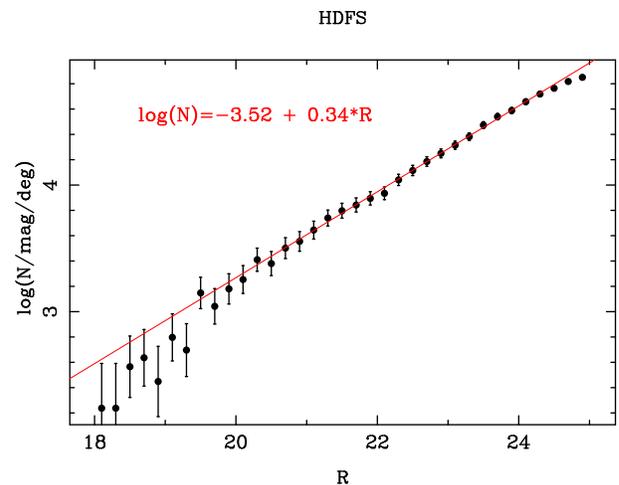}
\caption{
$R$-band galaxy counts per unit magnitude per square degree 
with Poisson error bars and solid fit of 
$\log_{10}(N) = -3.52 + 0.34 R$ based on counts 
at total AB magnitudes $20<R<24$.  
Incompleteness is minimal below total magnitude $R=24.5$ and 
reaches 50\% around $R=25.5$.  
\label{fig:Rcounts}
}
\end{figure}

  The sky density of objects in our catalog is greatly reduced near the brightest 
stars, leading us to create a mask to properly exclude these 
regions from analyses of sky density and clustering.  
Our methodology is described in detail in L.~Infante et al. 
(2005, in preparation).  
A careful analysis of the implied stellar and galaxy luminosity functions 
as a function of stellarity cut shows that stellar contamination is 
minimized with negligible loss of galaxy counts by requiring 
$stellarity < 0.8$.  
The total $R$-band magnitude 
counts (from AUTO) 
are shown in Figure \ref{fig:Rcounts} and appear 90\% complete 
to $R=24.5$ and 50\% complete to $R=25.5$.  
Part of the incompleteness comes from AUTO beginning to 
underestimate the flux of very dim objects, as noted by 
\citet{labbeetal03}.  
The fit for the number of galaxies per magnitude per square 
degree is $\log(N)=-3.52 + 0.34 R$.  Our slope of 0.34$\pm$0.01 agrees 
well with previous measurements of 0.321$\pm$0.001 by 
\citet{smailetal95}, 0.361$\pm$0.004 by \citet{capaketal04}, 
0.31$\pm$0.01 and 0.34$\pm$0.01 in two different fields by  
\citet{steidelh93}, and 0.39 by \citet{tyson88}.  Our differential 
counts at $R=25.5$ of 1.3$\times 10^5$ agree well with 
the values of 1.2$\times10^5$ reported by \citet{steidelh93}. 

\subsection{$UVR$
Photometric Selection of Lyman Break Galaxies}
\label{sec:lbg_select}

\begin{figure}[t!]
\epsscale{1.2}
\plotone{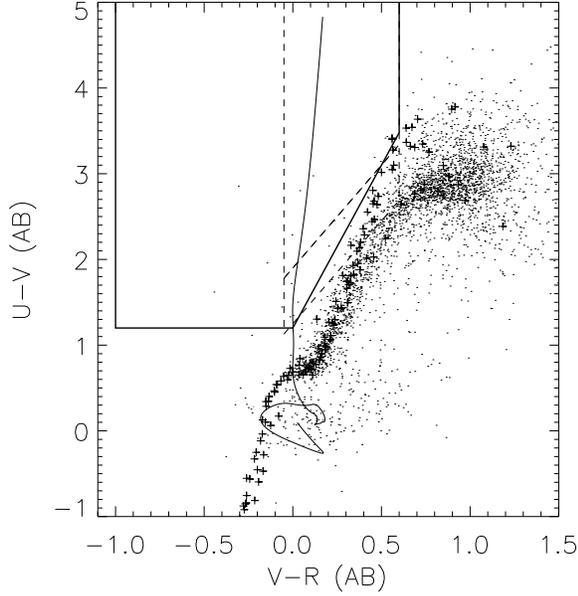}
\caption{
$U-V$ versus $V-R$ colors in our filter system (AB magnitudes) 
of a Lyman break galaxy 
template spectrum (Adelberger, priv. comm.) 
from redshift 0 to 3.4 (solid curve), the Pickles catalog 
of stars (plusses, \citealp{pickles98}), and the definite point sources in our 
catalog ($stellarity\ge 0.95$, points).  
 The C and MC color selection regions used by \citet{steideletal03} 
transformed into our UVR filter system are shown with dashed lines, 
and our adopted color selection region is shown with solid lines.  
\label{fig:uvr_stars}
}
\end{figure}

Various filter sets have yielded success at finding  
LBGs, including  
$U_n G {\cal R}$ \citep{steideletal96}, SDSS 
$u'g'r'i'z'$ \citep{bentzow04}, $u'BVRI$ \citep{cookeetal05},   
$G {\cal R} i$ \citep{steideletal99}, and 
$BRI$ (\citealt{gawiseretal01}, \citealt{prochaskaetal02}).  
Figure \ref{fig:uvr_stars} shows star colors expected and measured in 
our survey versus our adopted $UVR$ color selection 
criteria of
 \begin{eqnarray}
(U-V) &>& 1.2 \nonumber \\
 -1.0 &<& (V-R) \; \; < \; 0.6  \nonumber \\
(U-V) &>& 3.8 (V-R) \; + \; 1.2 \nonumber \\
19 &<& R \; \; \; < \; \; 25.5 \; \; \; , 
\label{eq:uvr}
\end{eqnarray}
where all magnitudes are AB. 
The color criteria shown as dashed lines in the figure result 
from the transformation of the 
$U_n$ $G$ ${\cal R}$ criteria described by \citet{steideletal99} 
into $UVR$.\footnote{For power-law spectra, the effective wavelengths of 
these filters imply the translations $(U-V)_{AB} = 1.25(U_n - G)$ and 
$(V-R)_{AB} = 0.50(G-{\cal R})$.}  
Although the upper branch of the stellar distribution at red $V-R$ 
colors represents giants which are rare at our survey depth and 
high galactic latitude, we shifted our criteria to avoid this region as it 
also contains dim dwarf stars with correspondingly large errors in color, 
which are the primary expected source of interlopers.  
The region of the 
Steidel et al. criteria avoided by this shift 
had the highest fraction of interlopers 
and also shows a high interloper fraction in the more 
liberal $UVR$-selected subset of the \citet{cookeetal05} LBG sample.   
The LBG template curve falls within our selection region for 
$3.0<z<3.4$.  
At higher redshifts, the LBGs rapidly become too red in 
$V-R$ to distinguish from lower-redshift objects and are better selected 
as dropouts in $B-R$.  
We allow the color selection region
 to extend far to the blue in $V-R$ to account for the 
effect of Lyman $\alpha$ emission lines, which add flux in $V$.

\begin{figure*}[t]
\plottwo{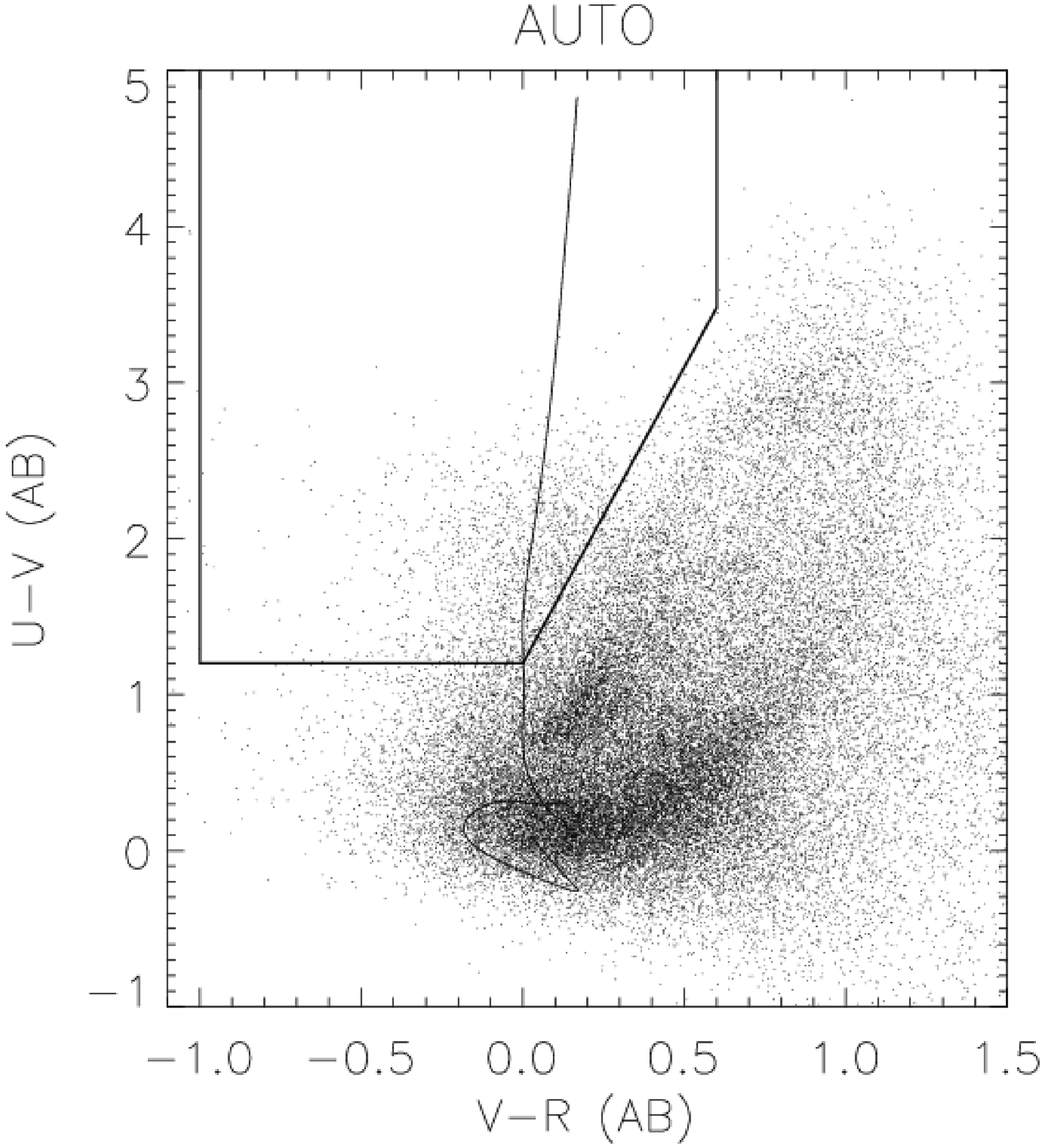}{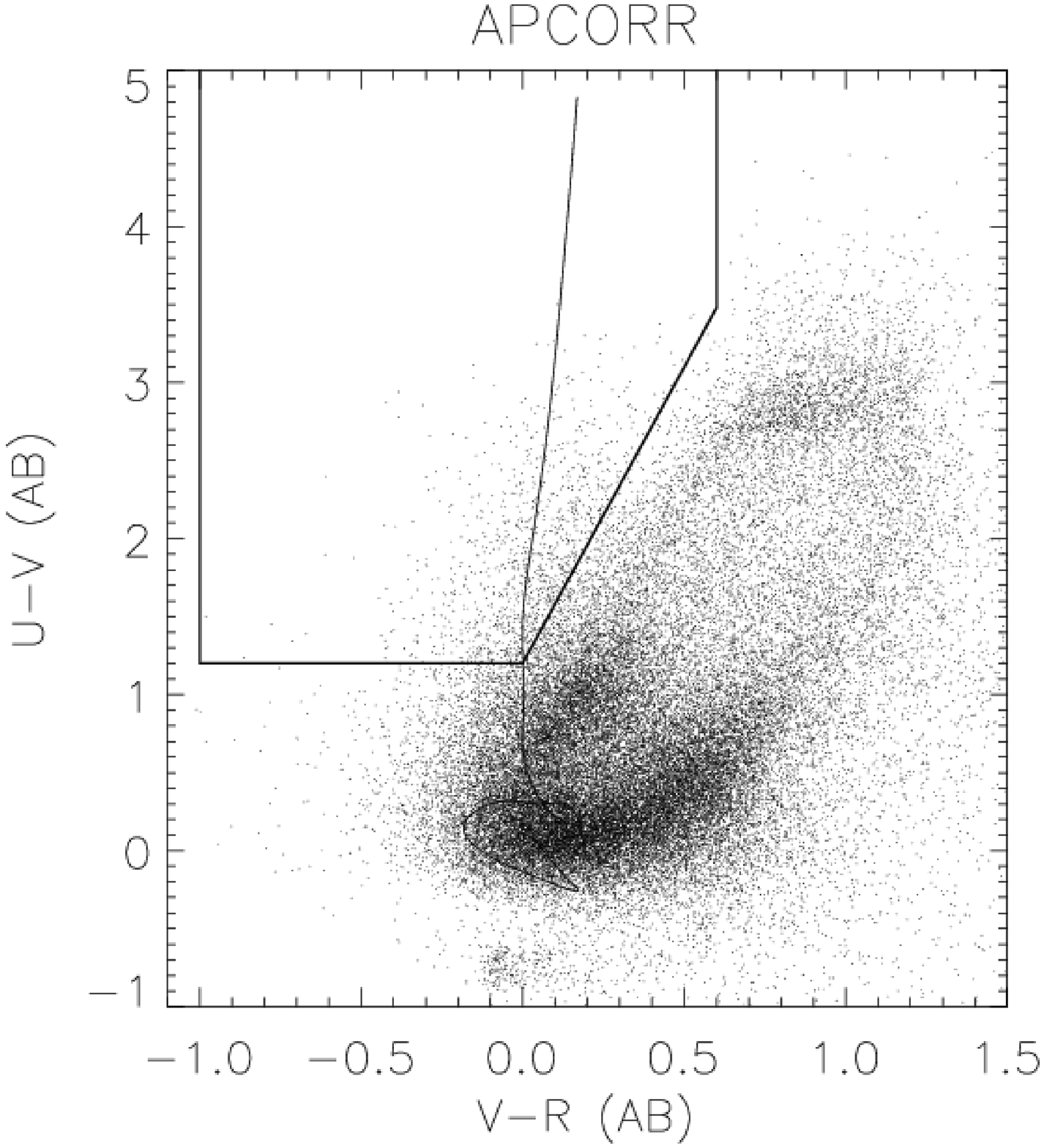}
\caption{
$U-V$ versus $V-R$ colors of our full catalog (points) 
in AB magnitudes from SExtractor AUTO (Kron 
elliptical aperture) photometry (left panel) and from APCORR (corrected 
aperture) photometry (right panel).
The LBG color selection region is shown with solid lines.  
The solid curve shows the colors of a 
Lyman break galaxy template spectrum over the interval $0<z<3.4$, with 
the interval $3.0<z<3.4$ falling in the selection region.
The reduced scatter reflects improved performance 
of the APCORR fluxes, with a clearer definition of the 
main sequence and M dwarf locus.    
\label{fig:uvr}
}
\end{figure*}

For spectroscopic selection and clustering analysis,
 objects with $U$ fluxes, $f_U$, less than their $1 \sigma$ flux uncertainty,  
$\sigma_U$,  
were assigned an upper limit of 
$f_U = \sigma_U$ to make it less likely that negative 
sky fluctuations were responsible for turning a dim object in $U$ 
into a formal dropout in $U-V$.  
This correction avoids lower-redshift 
interlopers at the cost of some incompleteness. 
The vast majority of our LBG candidates are formal dropouts in the $U$ filter, 
with $f_U < \sigma _U$ before being set to these limit values.  The limit 
values were used to generate the $U-V$ colors plotted in Fig. \ref{fig:uvr}, 
which shows our full catalog including 
$z\simeq3$ Lyman break galaxy candidates selected 
by their $UVR$ colors.

Our selection criteria yield 
1607 candidates in 1137 square arcminutes, or 1.4 arcmin$^{-2}$.  
\citet{steideletal99} found 
 1.2 arcmin$^{-2}$ at ${\cal R}<25.5$ using stricter criteria than ours, 
\citet{steideletal03} found 1.7  arcmin$^{-2}$ using the extended 
$U_n$ $G$ ${\cal R}$ criteria shown translated into $UVR$ in 
Figure \ref{fig:uvr_stars}, \citet{capaketal04} found 1.5 arcmin$^{-2}$ 
using UBR, 
and \citet{cookeetal05} found 1.5 arcmin$^{-2}$ 
using $u'VRI$ criteria.  
\citet{steideletal03} found a redshift distribution 
$<z>=2.96\pm0.29$.   Our $UVR$ criteria are expected to start 
selecting LBGs at redshifts higher by 0.16 because 
the red cutoff of the $U$-band filter and its 
effective wavelength after accounting for the MOSAIC II CCD QE 
are both $\sim200$\AA\ redder 
than for $U_n$.  Our criteria are expected to stop selecting LBGs 
at redshifts higher by up to 0.4 because the Lyman alpha forest 
causes red $V-R$ colors at a higher redshift than in $G-{\cal R}$ due to the 
longer effective wavelength of $V$.  Hence we expect $<z>=3.2\pm0.4$, 
and this will be calibrated via spectroscopy.

\begin{figure}[b!]
\epsscale{1.2}
\plotone{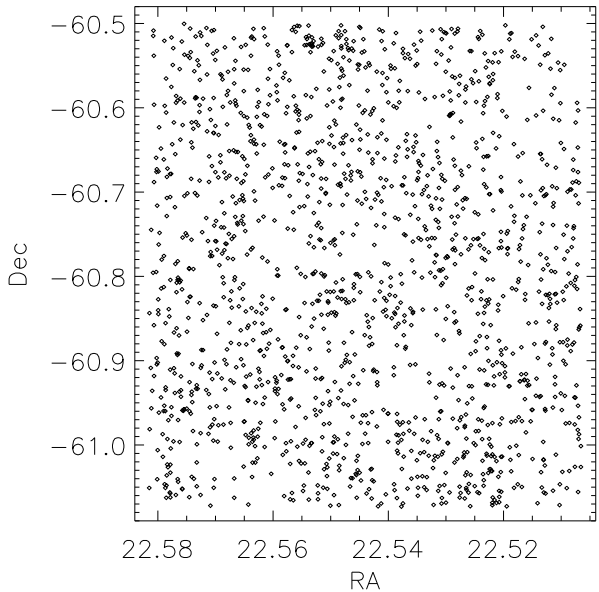}
\caption{
Sky distribution of the 1607 
LBG candidates 
in our catalog of E-HDFS 
showing some obvious clustering on small scales.  
\label{fig:lbg_radec}
}
\end{figure}

Figure \ref{fig:lbg_radec} shows the spatial distribution of these 
LBG candidates.  
We estimate the angular correlation function of these objects using 
the \citet{landys93} estimator 
\begin{equation}
w(\theta) = \frac{DD - 2DR + RR}{RR} \; \; \; ,
\label{eq:ls}
\end{equation}
where $DD$ represents the number of pairs at separation $\theta$ in 
the object catalog, $RR$ represents the appropriately-normalized 
number of pairs at separation $\theta$ in a Poisson-distributed 
mock catalog, and 
$DR$ represents the appropriately-normalized 
number of pairs at separation $\theta$ with one in the object catalog 
and the other in the mock catalog.  
This estimator has variance 
\begin{equation}
var[w(\theta)] \simeq {\frac{(1 + w(\theta))^2}{RR}} \; \; \; , 
\label{eq:ls_var}
\end{equation}
where the denominator gives the Poisson error contribution i.e. 
the number of pairs expected to be found in the 
bin centered on $\theta$ given the survey geometry and the 
total number of LBGs in the survey.\footnote{RR is equivalent to $1/p$ 
from \citealp{landys93}} 
The observed values of $w(\theta)$ must then be corrected for 
the integral constraint caused by estimating the true sky density 
of LBGs from our survey to find a power-law fit \citep{foucaudetal03}, 
\begin{equation}
A_w \theta^{- \beta} = w_{obs}(\theta) + 
\frac{1}{\Omega^2} \int \int w(\theta) d\Omega_1 d\Omega_2 
\; \; \; .
\label{eq:w_fit}
\end{equation}
In practice, this correction is performed by 
using a modified estimator 
\begin{equation}
w(\theta) = \frac{(1+A)DD - 2DR + RR}{RR} \; \; \; ,
\label{eq:ls-modified}
\end{equation}
where the correction factor $A$ is given by 
\begin{equation}
A = \frac{\sum w(\theta) RR}{\sum RR }  \; \; \; .
\label{eq:integral_constraint}
\end{equation}
A is estimated iteratively and usually converges in a few 
iterations.

\begin{figure}[t!]
\vspace{0.2in}
\includegraphics[angle=270,scale=0.40]{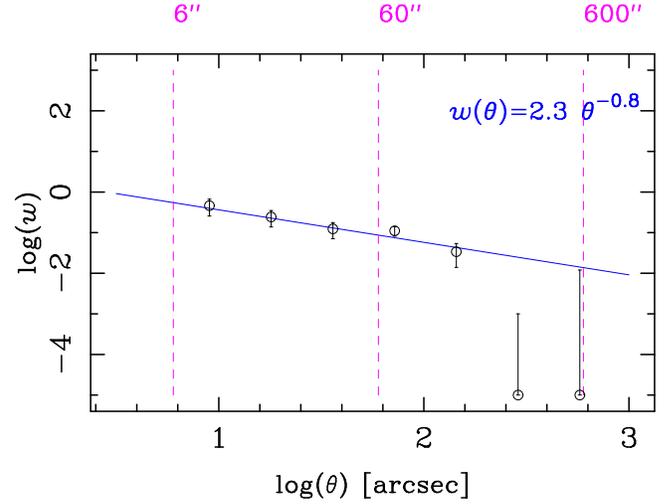}
\caption{
Angular auto-correlation function of UVR-selected LBG candidates (open circles with error bars) with fit (solid line) labeled.    
\label{fig:w_lbg}
}
\end{figure}

The angular correlation function found for Lyman break galaxy candidates 
in Figure \ref{fig:w_lbg} is $w(\theta) = (2.3\pm1.0) \theta^{-0.8}$, 
where $\theta$ is measured in arcsec, 
intermediate between previous results for fixed $\beta = 0.8$ 
from \citet{giavaliscoetal98} of $A_w = 1.4^{+1.5}_{-0.7}$ 
and \citet{foucaudetal03} of $A_w = 5.2 \pm 0.7$.  
This implies a correlation length for our sample of 
$r_0 = 5\pm1 $$h^{-1}_{70}$Mpc and an LBG bias factor of $b \simeq 2$
for the $\Lambda$CDM cosmology.  
These results  
will be updated after spectroscopic calibration of our $UVR$ 
LBG selection and by including our full square degree survey.

\section{CONCLUSIONS}
\label{sec:conclusions}

MUSYC is unique for its 
combination 
of depth and
total area, 
for the coverage in X-ray, UV, mid- and far-infrared
wavelengths and 
for providing the $UBVRIz'$ plus near-infrared photometry 
needed to produce high-quality photometric redshifts 
over a square-degree of sky.  
This multiwavelength coverage will enable comparison of selection 
effects which have previously complicated 
the study of galaxies in the 
high-redshift universe.  
The large area ensures good
statistics for studies of clustering, luminosity functions, and
surface densities.

We expect that the careful attention paid to optimized photometry 
methods and the estimation of accurate photometric uncertainties 
will allow our public data to make a significant impact for a 
wide range of scientific studies.  In particular, this high-quality 
$UBVRIz'$ data, combined with forthcoming near-infrared photometry, 
should yield photometric redshifts with good accuracy.  
It is interesting to note that some of the standard 
techniques perform well despite making a number of idealized 
assumptions.  For instance, we found that the 
optimal point source apertures are remarkably close in diameter 
to the idealized case of 1.35 fwhm due to a partial cancellation of the 
non-Gaussian PSF 
preferring a larger aperture and the correlated noise preferring 
a smaller aperture.  The more traditional approach of using $2''$ 
apertures produces biases in point-source colors.   
Alleviating those biases by ``PSF-matching'' i.e. smoothing the images to 
match the fwhm of the image with the worst seeing must be done carefully 
given the non-Gaussian PSF shapes and will result in degraded signal-to-noise 
for point sources. 
Estimates of point source detection depth that assume uncorrelated 
noise and extrapolate from the rms pixel noise will overestimate 
the depth by about 0.5 magnitudes for 2$''$ or 3$''$ apertures.  

The APCORR fluxes introduced in this paper
 perform extremely well, as seen in Fig. \ref{fig:uvr}.  
The nearly unbiased performance of APCORR fluxes for both point sources 
and extended objects is impressive, and the correlated errors 
between filters for extended objects 
are acceptable if the flux is misestimated by a constant 
fraction in all filters.  However, one should use caution when  
calculating photometric redshifts on extended objects in 
photometric catalogs generated using 
apertures 
which 
might miss a different fraction of the object flux
in each filter.    
The best results should be obtained by using 
the APCORR fluxes in our catalog for unresolved and barely resolved sources, 
with further investigation needed to determine whether APCORR outperforms 
AUTO for extended sources.  

Our $UBVRIz'$ images and catalogs of the $33'\times 34.5'$ Extended 
Hubble Deep Field South are available to the public 
at \url{http://www.astro.yale.edu/MUSYC} .

\acknowledgments

We thank Buell Januzzi and Frank Valdes for their detailed answers and 
trouble-shooting with questions on the MSCRED package and 
Benne W. Holwerda whose guide, ``Source Extractor for Dummies,'' 
provided a valuable supplement to the official SExtractor User's Manual 
written by E. Bertin.  
We acknowledge helpful conversations with 
Anton Koekemoer, David Spergel, and Nick Suntzeff.    
We thank the 
additional members of the 
MUSYC collaboration, including 
Danilo Marchesini, 
Ken Rines, 
Ned Taylor, and Bill van Altena
for their valuable input.      
We are grateful for 
support from Fundaci\'{o}n Andes, the FONDAP Centro de Astrof\'{\i}sica, 
and the Yale Perspectives in Science program.  
This material is based upon work supported by the National Science 
Foundation under Grant. No. AST-0201667, 
an NSF Astronomy and Astrophysics Postdoctoral 
Fellowship (AAPF) awarded to E.G.   
S.T. acknowledges the support of the Danish Natural Research Council.  
We thank the staff 
of Cerro Tololo Inter-American Observatory for their invaluable 
assistance with our observations.  
This research has made use of NASA's Astrophysics Data System.

\begin{appendix}

\section{Optimal Weights for Image Stacking for Point Sources and Extended Sources}
\label{sec:app-weights}

We wish to combine a series of processed individual images which have 
signal $S_i$ and noise $N_i$.  The exact signal obviously depends on which 
star is measured, but if a common set of stars are considered the variations 
in signal will be due to differences in exposure time, cirrus 
extinction, and atmospheric extinction.
The noise should be sky-dominated away from bright objects but 
measuring the rms in ``blank'' regions of sky properly accounts for the 
readnoise as well.
For surface-brightness-optimization we want to average 
the total signal from a set of stars, as measured by the IRAF 
routine {\it phot} or 
the inverse of the {\it mscscale} header output by {\it mscimatch}.      
For point-source-optimization we 
want to determine the average signal that falls within our eventual 
optimal aperture, which will have a diameter of roughly 
1.35 times the fwhm of the final image.  
The seeing in the final image can be estimated or measured from an 
initially unweighted stack, or an iterative procedure can be used.
Alternatively, a correction to the total star counts $S^{tot}_i$ 
can be calculated by integrating an assumed
 Gaussian PSF out to the diameter of 
the expected optimal aperture of radius R; if the   
seeing in each image is $\sigma_i =$ fwhm$_i/2.35$, 
we obtain 
\begin{equation}
S_i = S^{tot}_i [1 - \exp(-R^2/2 \sigma_i^2)] \; \; \; .
\label{eq:s_i}
\end{equation}
In either type of optimization, the noise value used should be the 
background rms for the expected aperture size, which depends on the level of 
noise correlation between pixels.  Assuming that the noise correlations 
have the same behavior in each individual image, it is sufficient 
to use the pixel-by-pixel rms i.e. $\sigma_1$ measured in each 
image since this will produce the right relative weights.

One can derive the weights that optimize the $S/N$ of the combined image 
by noting that the combined image will have $S=\sum_i w_i S_i$ and 
$N^2=\sum_i w_i^2 N_i^2$ (the weighted sum is just an unweighted 
sum of new images having signal $w_i S_i$ and noise $w_i N_i$).  
Maximizing the function $S/N$ 
produces the result $w_i = k S_i/N_i^2$, where  
multiplying all the weights by a constant $k$ preserves the 
final $S/N$.  
In astronomy, it is typical to first scale the images to have 
equal signal levels; this accounts for differences in exposure 
time and extinction to provide constant photometry and 
makes it easier to delete outlying pixels due to cosmic rays, satellite 
trails, etc.    
The IRAF-{\it mscred} routine 
{\it mscstack} performs this scaling by multiplying by the {\it mscscale}
header value and then 
uses a weightfile to weight the 
{\bf scaled} images.
So the weights used must be for the scaled images, which leads to the 
formula for surface-brightness-optimized weighting of
\begin{equation}
w_i^{SB}= \frac{1}{(mscscale_i \times rms_i)^2} \; \; \; ,
\end{equation}
which is equivalent 
to $(S_i/N_i)^2$ in the original image 
because the image being weighted is different than the 
one for which $S_i$ and $N_i$ were measured.  
The output image is the same since instead of multiplying 
$S_i$ by $S_i/N_i^2$ and adding we are equating the 
signal levels and then multiplying 1 by $(S_i/N_i)^2$, but 
performing the scaling first makes it easier to remove outlying 
pixels from this sum as mentioned above.  
The weights being used correspond to $S/N^2$ 
for the scaled images; they reduce to the familiar case of 
inverse-variance weighting since 
the signal levels have already been equalized.  
For point-source optimized weighting we use 
\begin{equation}
w_i^{PS}= \left(\frac{factor_i}{mscscale_i \times rms_i} \right) ^2 \; \; \; ,
\end{equation}
where $factor_i$ is given by 
\begin{equation}
factor_i = 1 - \exp \left( -1.3 \frac{fwhm_{stack}^2}{fwhm_i^2} \right)
\; \; \; ,
\end{equation}
where we have re-expressed the formula for $S_i$ from Eq. \ref{eq:s_i} 
in terms of the measured seeing in an unweighted (or 
surface-brightness optimized) stack, $fwhm_{stack}$, and that in 
each individual image $fwhm_i$.  Again, these weights are only valid 
when used on the post-scaled images, and they differ by a factor of 
$S_i$ from the weights appropriate for the original images.   

Due to the large number of individual exposures taken in e.g. 
$U$-band and multiple 
observing runs, it is sometimes necessary to perform stacking as an 
iterative procedure.  To do this, the signal, noise, and seeing 
in the intermediate stacked images should be measured empirically 
as before.
Then {\it mscstack} or the equivalent can 
be used to stack these intermediate images just as if they 
were individual images, scaling by the new {\it mscscale} values 
and calculating a new weights file using the above equations.  
As long as the final seeing is estimated well, no loss of 
signal-to-noise should occur from the iteration.  

A corollary question is when to discard an image with poor $S/N$ rather 
than to include it in the stack.  The formal answer appears to be never as 
$w_i > 0$ for all $i$.   
The magic of the optimal weighting formula can be 
seen:  
\begin{equation}
 \frac{S}{N} =  \frac{\sum_i w_i S_i}{\sqrt{\sum_i w_i^2 N_i^2}}
= \frac{\sum_i S_i^2/N_i^2}{\sqrt{\sum_i S_i^2/N_i^2}}
= \sqrt{\sum_i S_i^2/N_i^2} \; \; \; .
\end{equation}
Hence the optimal weights cause $S/N$ to add in quadrature so it will never 
formally decrease no matter how poor an input image is.  
Note that having a magnitude of extinction or twice 
the seeing still leaves an image with significant $S/N$; 
a $S/N$ 
threshold of 40\% would require discarding images with 
about 2 times worse 
seeing 
or 1 magnitude of extinction (if the sky background is unaffected).
The $S/N$ 
of the combination of the image with a given 
$S/N$ and an image with 40\% of that $S/N$
is 8\% higher than that of the better image alone so this 
is somewhat useful.  A conservative approach would be
 to cut entirely any images with $S/N < 10$\% that 
of the median image as a way of reducing systematic effects 
not accounted for by these idealized formulae.  

\section{Optimal Weights for Point Source Detection}
\label{sec:app-filter}

\citet{irwin85} showed using simulations that the best performance for 
point source detection was achieved by filtering the image with the 
PSF itself.  
This result is derived 
in the SExtractor manual
using 
Fourier transforms.
In practice due to the constraints of computation time one chooses to 
cut off the PSF at a finite radius, as is standard in the 
convolution kernels offered as part of the SExtractor package.  
The general case for an optimized weighted sum of pixel values
$S = \sum_j w_j S_j$, where each pixel has signal $S_j$ and 
noise $N_j$,  
can be derived in analogy to the result for adding images found above, 
yielding
$w_j \propto S_j/N_j^2$.  The constant background noise for 
sky-dominated objects thus leads to a filter of the precise shape of the 
PSF.  For objects significantly brighter than the sky, $S = N^2$ 
giving constant weight i.e. a tophat which gradually morphs into the  
roughly Gaussian shape of the PSF as you move far enough away from the 
object center for the sky to dominate the noise.  SExtractor 
only allows for a fixed convolution kernel so we optimized the detection 
for sky-dominated objects by using the PSF 
(truncated to 7$\times$7 pixels) 
as our filter.  
By the same argument, the optimum signal-to-noise measurement 
of the photometry of a point source will be obtained by a weighted sum 
of pixel fluxes weighted 
using the PSF shape itself centered at the barycenter of the object.  
This is typically referred to as ``PSF photometry'' but it is not 
a supported feature of SExtractor photometry, just of object 
detection.  Note that this is 
no longer strictly optimal when the noise is correlated between pixels 
as we have found in our images, since the derivation of the optimal 
weights for each pixel assumed uncorrelated noise.  
We decided against using PSF photometry for this reason, and also 
because many of our science objects are slightly 
resolved and this would overweight the fluxes in their core, 
increase the risk of color biases, and complicate the task 
of correcting the aperture fluxes of extended objects.

\end{appendix}

%

%

%
%

%

%

\begin{deluxetable}{llrllccc}
\tabletypesize{\small}
\tablecolumns{8}
\tablewidth{0pc}
\tablecaption{MUSYC Fields \label{tab:fields}
}
\tablehead{
\colhead{Field} & \colhead{RA} & \colhead{Dec} 
& \colhead{Galactic} & \colhead{Ecliptic} 
& \colhead{E(B-V)} & \colhead{100$\mu$m Em.} & \colhead{N(HI)}
\\ 
\colhead{}      & \colhead{}           & \colhead{}  
& \colhead{Coords. (deg)} & \colhead{Coords. (deg)} 
& \colhead{} & \colhead{(MJy/sr)}  & \colhead{(cm$^{-2}$)} 
}
\startdata
E-HDFS & 22:32:35.6 & -60:47:12 & (328,-49)& (311,-47)&0.03& 1.37& 1.6E+20\\
E-CDFS & 03:32:29.0 & -27:48:47 & (224,-54)&(41,-45)&0.01& 0.40& 9.0E+19\\
SDSS1030+05 & 10:30:27.1&05:24:55& (239,50)&(157,-4)&0.02&1.01&2.3E+20\\
CW1255+01 & 12:55:40&01:07:00&    (306,64)& (192,7)& 0.02& 0.81& 1.6E+20\\
\enddata
\end{deluxetable}

\begin{deluxetable}{lcrrc}
\tablecolumns{4}
\tablewidth{0pc}
\tablecaption{Optical Observations of E-HDFS with CTIO4m+MOSAIC-II
\label{tab:obs}
}
\tablehead{
\colhead{Dates} & \colhead{Filter} & \colhead{Number of} & \colhead{Exposure} & \colhead{Seeing} 
\\
\colhead{}     & \colhead{}        & \colhead{Exposures} & \colhead{Time (s)}  & \colhead{($''$)} 
}
\startdata
2002 Oct 6,8    & $U$ &47 &28200 & 1.40 \\
2002 Oct 6,8,10 & $B$ &13 &7500 & 1.35 \\
2002 Oct 6,10,12& $V$ &30 &10440 & 0.90 \\
2002 Oct 6,10,12& $R$ &21 &6300 & 0.85 \\
2002 Oct 6,12   & $I$ &26 &6300 & 0.85 \\
2002 Oct 6,12   & $z'$ &11 &2700 & 0.90 \\
2003 May 26,27  & $U$ &25 &15000& 1.20 \\
2003 May 27     & $B$ &10 &6000 & 1.10 \\
2003 May 27,28  & $R$ &20 &6000 & 1.10 \\
2003 May 26     & $z'$ &15 &3600 & 1.30 \\
\enddata
\end{deluxetable}

\begin{deluxetable}{ccclcr}
\tablecolumns{4}
\tablewidth{0pc}
\tablecaption{Results of photometric calibration for 2002 October 6 
in AB magnitudes\label{tab:photcal}
}
\tablehead{
\colhead{Filter} & \colhead{Airmass} & \colhead{Seeing} & \colhead{Zeropoint} 
& \colhead{Airmass coeff.} & \colhead{Color term}\\
\colhead{}     & \colhead{} & \colhead{($''$)} & \colhead{} & \colhead{} 
& \colhead{} 
}
\startdata
U& 1.34& 2.1& 24.15$\pm$0.012& 0.55& (-0.08$\pm$0.01)$\times(U-B)_{\mathrm{AB}}$\\
B& 1.38& 1.8& 25.24$\pm$0.009& 0.24& (-0.03$\pm$0.004)$\times(U-V)_{\mathrm{AB}}$\\
V& 1.40& 1.6& 25.64$\pm$0.004& 0.15& (0.03$\pm$0.003)$\times(B-R)_{\mathrm{AB}}$\\
R& 1.42& 1.5& 25.95$\pm$0.005& 0.09& (-0.009$\pm$0.005)$\times(V-I)_{\mathrm{AB}}$\\
I& 1.44& 1.5& 25.40$\pm$0.006& 0.05& (0.006$\pm$0.017)$\times(R-I)_{\mathrm{AB}}$ \\
$z'$& 1.46& 1.5& 24.61$\pm$0.014& 0.03& (-0.06$\pm$0.054)$\times(R-I)_{\mathrm{AB}}$\\
\enddata
\end{deluxetable}

\begin{deluxetable}{crccc}
\tablecolumns{6}
\tablewidth{0pc}
\tablecaption{Total Exposure Time and Photometric Zeropoints for E-HDFS Images\tablenotemark{a} \label{tab:final}
 }
\tablehead{
\colhead{Filter} & \colhead{Exposure} 
& \colhead{Mag. Zeropoint} 
& \colhead{Flux Zeropoint} 
\\
\colhead{}     & \colhead{Time (s)}     
&\colhead{(AB mag @ 1 count s$^{-1}$)} 
& \colhead{($\mu$Jy (count/s)$^{-1}$)} 
}
\startdata
$BVR$ &35340    &24.29& 0.698                 \\
U   &42600     &23.35& 1.660                  \\
B   &13200     &24.85& 0.417                  \\
V   &10440     &25.40& 0.251                  \\
R   & 11700    &25.75& 0.182                  \\
I   & 6000    &25.27& 0.283                  \\
$z'$   & 6000      &24.39& 0.637                 \\
\enddata
\tablenotetext{a}{before correction for E(B-V)=0.03}
\end{deluxetable}

\begin{deluxetable}{clll}
\tablecolumns{4}
\tablewidth{0pc}
\tablecaption{Fit parameters for background fluctuations in counts $s^{-1}$
as a function of aperture size using Eq. \ref{eq:power}\label{tab:noise}
 }
\tablehead{
\colhead{Filter} & \colhead{$\sigma_1$} & \colhead{$\alpha$}&\colhead{$\beta$} 
}
\startdata
BVR&    0.0029&     0.68&   1.47\\
U&      0.0012&     0.82&   1.30\\
B&      0.0043&     0.88&   1.33\\
V&      0.014 &     0.69&   1.34\\
R&      0.019 &     0.69&   1.40\\
I&      0.032 &     0.72&   1.42\\
$z'$&      0.030 &     0.87&   1.41\\	
\enddata
\end{deluxetable}

\begin{deluxetable}{ccccccc}
\tabletypesize{\small}
\tablecolumns{7}
\tablewidth{0pc}
\tablecaption{Seeing, Apertures, and Source Detection Limits for E-HDFS Images
\label{tab:apertures}}
\tablehead{
\colhead{Filter} & \colhead{IRAF fwhm} 
& \colhead{SE fwhm}&\colhead{SE $r_{1/2}$}
& \colhead{Optimal Aperture} 
& \colhead{Flux Enclosed} 
& \colhead{Source Detection}
\\
\colhead{}     & \colhead{Mode ($''$)} & \colhead{Median ($''$)} 
& \colhead{Median ($''$)}
& \colhead{($''$)} &\colhead{(Fractional)} & \colhead{Limit (5 $\sigma$, AB)} 
}
\startdata
BVR & 0.95   & 0.99   & 0.59 & 1.2 & 0.48 & 26.3\\     
U   & 1.30     & 1.48   & 0.78 & 1.6 & 0.50 & 26.0\\   
B   & 1.25     & 1.29   & 0.71 & 1.4 & 0.47 & 26.1\\   
V   & 0.90     & 0.96   & 0.56 & 1.2 & 0.52 & 26.0\\   
R   & 0.90     & 0.97   & 0.58 & 1.2 & 0.49 & 25.8\\   
I   & 0.80     & 0.99   & 0.56 & 1.2 & 0.49 & 24.7\\   
$z'$   & 0.90     & 1.06   & 0.62 & 1.2 & 0.43 & 23.6\\
\enddata
\end{deluxetable}

\begin{deluxetable}{lllllllll}
\tablecolumns{9}
\tabletypesize{\tiny}
\tablewidth{0pc}
\tablecaption{Optical source catalog: astrometric and photometric parameters 
determined from $BVR$ detection image and $UBVRIz'$ photometry.\label{tab:catalog}
}
\tablehead{
\colhead{Number} & \colhead{Name} & \colhead{RA} & \colhead{Dec} 
& \colhead{x} &\colhead{y} & \colhead{Stellarity} 
& \colhead{r$_{1/2}$} 
& \colhead{A\_IMAGE}\\
\colhead{}& \colhead{MUSYC-} & \colhead{(hrs)} & \colhead{(deg)}  
&\colhead{} &\colhead{} & \colhead{} 
& \colhead{($''$)} &\colhead{($''$)}\\ 
\colhead{B\_IMAGE} & \colhead{$\theta$\_IMAGE}
&\colhead{FLAGS}&
\colhead{$f$\_auto\_U} & \colhead{$\sigma$\_auto\_U} 
& \colhead{$f$\_auto\_B} & \colhead{$\sigma$\_auto\_B} 
& \colhead{$f$\_auto\_V} & \colhead{$\sigma$\_auto\_V}\\ 
\colhead{($''$)} & \colhead{ccw from X-axis}
&\colhead{} & 
\colhead{($\mu$Jy)} & \colhead{($\mu$Jy)}  
& \colhead{($\mu$Jy)} & \colhead{($\mu$Jy)}  
& \colhead{($\mu$Jy)} & \colhead{($\mu$Jy)}\\  
\colhead{$f$\_auto\_R} & \colhead{$\sigma$\_auto\_R} 
& \colhead{$f$\_auto\_I} & \colhead{$\sigma$\_auto\_I} 
& \colhead{$f$\_auto\_$z'$} & \colhead{$\sigma$\_auto\_$z'$} &
\colhead{$f$\_apcorr\_U} & \colhead{$\sigma$\_apcorr\_U} 
& \colhead{$f$\_apcorr\_B}\\
 \colhead{($\mu$Jy)} & \colhead{($\mu$Jy)}  
& \colhead{($\mu$Jy)} & \colhead{($\mu$Jy)}  
& \colhead{($\mu$Jy)} & \colhead{($\mu$Jy)}  
& \colhead{($\mu$Jy)} & \colhead{($\mu$Jy)}  
& \colhead{($\mu$Jy)}\\
\colhead{$\sigma$\_apcorr\_B} 
& \colhead{$f$\_apcorr\_V} & \colhead{$\sigma$\_apcorr\_V} 
& \colhead{$f$\_apcorr\_R} & \colhead{$\sigma$\_apcorr\_R} 
& \colhead{$f$\_apcorr\_I} & \colhead{$\sigma$\_apcorr\_I} 
& \colhead{$f$\_apcorr\_$z'$} & \colhead{$\sigma$\_apcorr\_$z'$}\\
 \colhead{($\mu$Jy)}  
& \colhead{($\mu$Jy)} & \colhead{($\mu$Jy)}  
& \colhead{($\mu$Jy)} & \colhead{($\mu$Jy)}  
& \colhead{($\mu$Jy)} & \colhead{($\mu$Jy)}  
& \colhead{($\mu$Jy)} & \colhead{($\mu$Jy)}\\
}
\startdata
 0&J223023.68-604817.4&22.506580&-60.804850&7397.04&3613.83&0.490000&0.377004&0.0771630\\
0.0771630&45.0000&16&
0.00599998&0.0123559&0.0268263&0.0133826&0.0721889&0.0189753\\
0.0535715&0.0192685&0.0584936&0.0536170&0.105203&0.142492&
0.0153194&0.0299205&0.0582284\\
0.0300672&0.129481&0.0322477&0.0994542&0.0346933&0.0963358&0.0964758&0.129852&0.292286\\
 1&J223024.28-604008.9&22.506747&-60.669150&7395.69&5443.56&0.380000&0.381543&0.0771630\\
0.0771630&45.0000&16&
0.0545032&0.0134836&0.0528900&0.0133148&0.0600472&0.0181627\\
0.0531777&0.0189421&0.0346594&0.0537312&-0.0605122&0.0999073&
0.123143&0.0319542&0.111140\\
0.0292123&0.0983857&0.0301251&0.0985063&0.0331430&0.0733320&0.0938256&-0.113113&0.287106\\
 2&J223023.85-604558.3&22.506627&-60.766210&7397.14&4134.88&0.490000&0.555360&0.216003\\
0.133500&-0.200000&24&
0.0909909&0.0243018&0.0549541&0.0255182&0.0440691&0.0353969\\
0.130563&0.0386372&0.159939&0.108677&0.0605380&0.286208&
0.100151&0.0285476&0.0699377\\
0.0270144&0.0861040&0.0280836&0.125653&0.0311628&0.113170&0.0863858&0.0282574&0.261025\\
 3&J223024.98-603027.9&22.506940&-60.507770&7394.68&7619.57&0.940000&0.328410&0.199716\\
0.120951&18.8000&24&
-0.0217973&0.0159112&0.000543353&0.0123818&0.115021&0.0169463\\
300.840&0.0511371&577.641&0.128364&-0.00643368&0.124763&
-0.0458961&0.0422158&-0.000965204\\
0.0289711&0.222445&0.0300914&601.662&0.125434&1191.94&0.301193&0.00430643&0.280111\\
 4&J223023.35-605242.2&22.506487&-60.878400&7397.96&2622.20&0.610000&0.613032&0.369528\\
0.202653&-5.40000&24&
-0.00917560&0.0288420&0.0469781&0.0383380&0.134234&0.0542395\\
0.155328&0.0622344&-0.104166&0.130648&0.110060&0.458585&
-0.0224367&0.0274903&0.100572\\
0.0275703&0.155969&0.0290837&0.143652&0.0332477&0.0475046&0.0869854&0.677608&0.276257\\
\enddata
\tablecomments{This table is published in its entirety in the electronic edition of the Astrophysical Journal. A portion is shown here for guidance regarding its form and content. Due to geometry, these first few objects are near the image 
border and have poor photometry, leading to their large FLAG values. }
\end{deluxetable}


\begin{thebibliography}{58}
\expandafter\ifx\csname natexlab\endcsname\relax\def\natexlab#1{#1}\fi

\bibitem[{{Abazajian} {et~al.}(2005)}]{abazajianetal05}
{Abazajian}, K. {et~al.} 2005, \aj, 129, 1755 

\bibitem[{{Altmann} {et~al.}(2004){Altmann}, {M\'{e}ndez}, {Ruiz}, {van
  Altena}, {Gawiser}, \& {van Dokkum}}]{altmannetal04}
{Altmann}, M., {M\'{e}ndez}, R.~A., {Ruiz}, M.~T., {van Altena}, W., {Gawiser},
  E., \& {van Dokkum}, P. 2004, in 14th European Workshop on White Dwarfs, ASP
  Conference Series, ed. D.~{Koester} \& S.~{Moehler}, Vol. 999

\bibitem[{{Arnouts} {et~al.}(2001)}]{arnoutsetal01}
{Arnouts}, S. {et~al.} 2001, \aap, 379, 740

\bibitem[{{Bechtold} {et~al.}(2003)}]{bechtoldetal03}
{Bechtold}, J. {et~al.} 2003, \apj, 588, 119

\bibitem[{{Bentz} {et~al.}(2004){Bentz}, {Osmer}, \& {Weinberg}}]{bentzow04}
{Bentz}, M.~C., {Osmer}, P.~S., \& {Weinberg}, D.~H. 2004, \apjl, 600, L19

\bibitem[{{Bertin} \& {Arnouts}(1996)}]{bertina96}
{Bertin}, E. \& {Arnouts}, S. 1996, \aaps, 117, 393

\bibitem[{{Burstein} \& {Heiles}(1978)}]{bursteinh78}
{Burstein}, D. \& {Heiles}, C. 1978, \apj, 225, 40

\bibitem[{{Capak} {et~al.}(2004)}]{capaketal04}
{Capak}, P. {et~al.} 2004, \aj, 127, 180

\bibitem[{{Chen} {et~al.}(2002)}]{chenetal02}
{Chen}, H. {et~al.} 2002, \apj, 570, 54

\bibitem[{{Cooke} {et~al.}(2005){Cooke}, {Wolfe}, {Prochaska}, \&
  {Gawiser}}]{cookeetal05}
{Cooke}, J., {Wolfe}, A.~M., {Prochaska}, J.~X., \& {Gawiser}, E. 2005, \apj,
  621, 596

\bibitem[{{Daddi} {et~al.}(2004){Daddi}, {Cimatti}, {Renzini}, {Vernet},
  {Conselice}, {Pozzetti}, {Mignoli}, {Tozzi}, {Broadhurst}, {di Serego
  Alighieri}, {Fontana}, {Nonino}, {Rosati}, \& {Zamorani}}]{daddietal04a}
{Daddi}, E., {Cimatti}, A., {Renzini}, A., {Vernet}, J., {Conselice}, C.,
  {Pozzetti}, L., {Mignoli}, M., {Tozzi}, P., {Broadhurst}, T., {di Serego
  Alighieri}, S., {Fontana}, A., {Nonino}, M., {Rosati}, P., \& {Zamorani}, G.
  2004, \apjl, 600, L127

\bibitem[{{Davis} {et~al.}(2003)}]{davisetal03}
{Davis}, M. {et~al.} 2003, in Discoveries and Research Prospects from 6- to
  10-Meter-Class Telescopes II. Edited by Guhathakurta, Puragra. Proceedings of
  the SPIE, Volume 4834, 161

\bibitem[{{Dickinson} {et~al.}(2004)}]{dickinsonetal04}
{Dickinson}, M. {et~al.} 2004, \apjl, 600, L99

\bibitem[{{Farrah} {et~al.}(2004){Farrah}, {Priddey}, {Wilman}, {Haehnelt}, \&
  {McMahon}}]{farrahetal04}
{Farrah}, D., {Priddey}, R., {Wilman}, R., {Haehnelt}, M., \& {McMahon}, R.
  2004, \apjl, 611, L13

\bibitem[{{Foucaud} {et~al.}(2003){Foucaud}, {McCracken}, {Le F{\` e}vre},
  {Arnouts}, {Brodwin}, {Lilly}, {Crampton}, \& {Mellier}}]{foucaudetal03}
{Foucaud}, S., {McCracken}, H.~J., {Le F{\` e}vre}, O., {Arnouts}, S.,
  {Brodwin}, M., {Lilly}, S.~J., {Crampton}, D., \& {Mellier}, Y. 2003, \aap,
  409, 835

\bibitem[{{Franx} {et~al.}(2003)}]{franxetal03}
{Franx}, M. {et~al.} 2003, \apjl, 587, L79

\bibitem[{{Fukugita} {et~al.}(1996){Fukugita}, {Ichikawa}, {Gunn}, {Doi},
  {Shimasaku}, \& {Schneider}}]{fukugitaetal96}
{Fukugita}, M., {Ichikawa}, T., {Gunn}, J.~E., {Doi}, M., {Shimasaku}, K., \&
  {Schneider}, D.~P. 1996, \aj, 111, 1748

\bibitem[{{Gawiser} {et~al.}(2001){Gawiser}, {Wolfe}, {Prochaska}, {Lanzetta},
  {Yahata}, \& {Quirrenbach}}]{gawiseretal01}
{Gawiser}, E., {Wolfe}, A.~M., {Prochaska}, J.~X., {Lanzetta}, K.~M., {Yahata},
  N., \& {Quirrenbach}, A. 2001, \apj, 562, 628

\bibitem[{{Giacconi} {et~al.}(2002)}]{giacconietal02}
{Giacconi}, R. {et~al.} 2002, \apjs, 139, 369

\bibitem[{{Giavalisco} {et~al.}(1998){Giavalisco}, {Steidel}, {Adelberger},
  {Dickinson}, {Pettini}, \& {Kellogg}}]{giavaliscoetal98}
{Giavalisco}, M., {Steidel}, C.~C., {Adelberger}, K.~L., {Dickinson}, M.~E.,
  {Pettini}, M., \& {Kellogg}, M. 1998, \apj, 503, 543

\bibitem[{{Giavalisco} {et~al.}(2004)}]{giavaliscoetal04}
{Giavalisco}, M. {et~al.} 2004, \apjl, 600, L93

\bibitem[{{Irwin}(1985)}]{irwin85}
{Irwin}, M.~J. 1985, \mnras, 214, 575

\bibitem[{{Jannuzi} \& {Dey}(1999)}]{jannuzid99}
{Jannuzi}, B.~T. \& {Dey}, A. 1999, in Astronomical Society of the Pacific
  Conference Series, 111

\bibitem[{{Labb{\' e}} {et~al.}(2003)}]{labbeetal03}
{Labb{\' e}}, I. {et~al.} 2003, \aj, 125, 1107

\bibitem[{{Lacy} {et~al.}(2004)}]{lacyetal04}
{Lacy}, M. {et~al.} 2004, \apjs, 154, 166

\bibitem[{{Landolt}(1992)}]{landolt92}
{Landolt}, A.~U. 1992, \aj, 104, 340

\bibitem[{{Landy} \& {Szalay}(1993)}]{landys93}
{Landy}, S.~D. \& {Szalay}, A.~S. 1993, \apj, 412, 64

\bibitem[{{Le F{\` e}vre} {et~al.}(2004)}]{lefevreetal04}
{Le F{\` e}vre}, O. {et~al.} 2004, \aap, 417, 839

\bibitem[{{Lira} {et~al.}(2004)}]{liraetal04}
{Lira}, P. {et~al.} 2004, ArXiv Astrophysics e-prints, astro-ph/0407396

\bibitem[{{MacDonald} {et~al.}(2004){MacDonald}, {Allen}, {Dalton},
  {Moustakas}, {Heymans}, {Edmondson}, {Blake}, {Clewley}, {Hammell}, {Olding},
  {Miller}, {Rawlings}, {Wall}, {Wegner}, \& {Wolf}}]{macdonaldetal04}
{MacDonald}, E.~C., {Allen}, P., {Dalton}, G., {Moustakas}, L.~A., {Heymans},
  C., {Edmondson}, E., {Blake}, C., {Clewley}, L., {Hammell}, M.~C., {Olding},
  E., {Miller}, L., {Rawlings}, S., {Wall}, J., {Wegner}, G., \& {Wolf}, C.
  2004, \mnras, 352, 1255

\bibitem[{{Palunas} {et~al.}(2000)}]{palunasetal00}
{Palunas}, P. {et~al.} 2000, \apj, 541, 61

\bibitem[{{Pickles}(1998)}]{pickles98}
{Pickles}, A.~J. 1998, \pasp, 110, 863

\bibitem[{{Prochaska} {et~al.}(2002){Prochaska}, {Gawiser}, {Wolfe},
  {Quirrenbach}, {Lanzetta}, {Chen}, {Cooke}, \& {Yahata}}]{prochaskaetal02}
{Prochaska}, J.~X., {Gawiser}, E., {Wolfe}, A.~M., {Quirrenbach}, A.,
  {Lanzetta}, K.~M., {Chen}, H., {Cooke}, J., \& {Yahata}, N. 2002, \aj, 123,
  2206

\bibitem[{{Radovich} {et~al.}(2004)}]{radovichetal04}
{Radovich}, M. {et~al.} 2004, \aap, 417, 51

\bibitem[{{Rix} {et~al.}(2004)}]{rixetal04}
{Rix}, H. {et~al.} 2004, \apjs, 152, 163

\bibitem[{{Rosati} {et~al.}(2002)}]{rosatietal02}
{Rosati}, P. {et~al.} 2002, \apj, 566, 667

\bibitem[{{Schlegel} {et~al.}(1998){Schlegel}, {Finkbeiner}, \&
  {Davis}}]{schlegelfd98}
{Schlegel}, D.~J., {Finkbeiner}, D.~P., \& {Davis}, M. 1998, \apj, 500, 525

\bibitem[{{Smail} {et~al.}(1995){Smail}, {Hogg}, {Yan}, \&
  {Cohen}}]{smailetal95}
{Smail}, I., {Hogg}, D.~W., {Yan}, L., \& {Cohen}, J. 1995, \apjl, 449, L105

\bibitem[{{Smith} {et~al.}(2002)}]{smithetal02}
{Smith}, J.~A. {et~al.} 2002, \aj, 123, 2121

\bibitem[{{Steidel} {et~al.}(1999){Steidel}, {Adelberger}, {Giavalisco},
  {Dickinson}, \& {Pettini}}]{steideletal99}
{Steidel}, C.~C., {Adelberger}, K.~L., {Giavalisco}, M., {Dickinson}, M., \&
  {Pettini}, M. 1999, \apj, 519, 1

\bibitem[{{Steidel} {et~al.}(2003){Steidel}, {Adelberger}, {Shapley},
  {Pettini}, {Dickinson}, \& {Giavalisco}}]{steideletal03}
{Steidel}, C.~C., {Adelberger}, K.~L., {Shapley}, A.~E., {Pettini}, M.,
  {Dickinson}, M., \& {Giavalisco}, M. 2003, \apj, 592, 728

\bibitem[{{Steidel} {et~al.}(1996{\natexlab{a}}){Steidel}, {Giavalisco},
  {Dickinson}, \& {Adelberger}}]{steideletal96b}
{Steidel}, C.~C., {Giavalisco}, M., {Dickinson}, M., \& {Adelberger}, K.~L.
  1996{\natexlab{a}}, \aj, 112, 352

\bibitem[{{Steidel} {et~al.}(1996{\natexlab{b}}){Steidel}, {Giavalisco},
  {Pettini}, {Dickinson}, \& {Adelberger}}]{steideletal96}
{Steidel}, C.~C., {Giavalisco}, M., {Pettini}, M., {Dickinson}, M., \&
  {Adelberger}, K.~L. 1996{\natexlab{b}}, \apjl, 462, L17

\bibitem[{{Steidel} \& {Hamilton}(1993)}]{steidelh93}
{Steidel}, C.~C. \& {Hamilton}, D. 1993, \aj, 105, 2017

\bibitem[{{Steidel} {et~al.}(2002){Steidel}, {Hunt}, {Shapley}, {Adelberger},
  {Pettini}, {Dickinson}, \& {Giavalisco}}]{steideletal02}
{Steidel}, C.~C., {Hunt}, M.~P., {Shapley}, A.~E., {Adelberger}, K.~L.,
  {Pettini}, M., {Dickinson}, M., \& {Giavalisco}, M. 2002, \apj, 576, 653

\bibitem[{{Szalay} {et~al.}(1999){Szalay}, {Connolly}, \&
  {Szokoly}}]{szalaycs99}
{Szalay}, A.~S., {Connolly}, A.~J., \& {Szokoly}, G.~P. 1999, \aj, 117, 68

\bibitem[{{Teplitz} {et~al.}(2001){Teplitz}, {Hill}, {Malumuth}, {Collins},
  {Gardner}, {Palunas}, \& {Woodgate}}]{teplitzetal01}
{Teplitz}, H.~I., {Hill}, R.~S., {Malumuth}, E.~M., {Collins}, N.~R.,
  {Gardner}, J.~P., {Palunas}, P., \& {Woodgate}, B.~E. 2001, \apj, 548, 127

\bibitem[{{Treister} {et~al.}(2004)}]{treisteretal04b}
{Treister}, E. {et~al.} 2004, \apj, 616, 123

\bibitem[{{Tyson}(1988)}]{tyson88}
{Tyson}, J.~A. 1988, \aj, 96, 1

\bibitem[{{van Dokkum} {et~al.}(2003)}]{vandokkumetal03}
{van Dokkum}, P.~G. {et~al.} 2003, \apjl, 587, L83

\bibitem[{{van Dokkum} {et~al.}(2004)}]{vandokkumetal04}
---. 2004, \apj, 611, 703

\bibitem[{{Williams} {et~al.}(1996)}]{williamsetal96}
{Williams}, R.~E. {et~al.} 1996, \aj, 112, 1335

\bibitem[{{Williams} {et~al.}(2000)}]{williamsetal00}
---. 2000, \aj, 120, 2735

\bibitem[{{Wolf} {et~al.}(2001){Wolf}, {Dye}, {Kleinheinrich}, {Meisenheimer},
  {Rix}, \& {Wisotzki}}]{wolfetal01}
{Wolf}, C., {Dye}, S., {Kleinheinrich}, M., {Meisenheimer}, K., {Rix}, H.-W.,
  \& {Wisotzki}, L. 2001, \aap, 377, 442

\bibitem[{{Wolf} {et~al.}(2004)}]{wolfetal04}
{Wolf}, C. {et~al.} 2004, \aap, 421, 913

\bibitem[{{Wolfe} {et~al.}(2005){Wolfe}, {Gawiser}, \& {Prochaska}}]{wolfegp05}
{Wolfe}, A.~M., {Gawiser}, E., \& {Prochaska}, J.~X. 2005, \araa, 43, 861

\bibitem[{{Wolfe} {et~al.}(1986){Wolfe}, {Turnshek}, {Smith}, \&
  {Cohen}}]{wolfeetal86}
{Wolfe}, A.~M., {Turnshek}, D.~A., {Smith}, H.~E., \& {Cohen}, R.~D. 1986,
  \apjs, 61, 249

\bibitem[{{York} {et~al.}(2000)}]{yorketal00}
{York}, D.~G. {et~al.} 2000, \aj, 120, 1579

\end{thebibliography}
\end{document}